\documentclass{aa}  
\usepackage{amsmath,amssymb}
\usepackage{chemformula}
\usepackage{txfonts}
\usepackage{threeparttable}
\usepackage{longtable}
\usepackage{booktabs}
\usepackage{ltablex}
\usepackage{caption}
\usepackage{hyperref}
\usepackage{xcolor}
\usepackage{booktabs}
\usepackage{siunitx}
\usepackage{utfsym} 

\begin{document}

     \title{Expanding the C$_3$H$_6$O$_2$ Isomeric Interstellar Inventory: Discovery of Lactaldehyde and Methoxyacetaldehyde in G+0.693-0.027}
    \authorrunning{Sanz-Novo et al.}
    \titlerunning{Expanding the C$_3$H$_6$O$_2$ Isomeric Interstellar Inventory}
    
    \author{M. Sanz-Novo, \inst{1}
           V. M. Rivilla, \inst{1}
           I. Jim\'enez-Serra, \inst{1}
           L. Colzi, \inst{1}
           S. Zeng,\inst{2}
           A. Meg\'ias, \inst{1}
           D. San Andr\'es, \inst{1,3}
           \'A. L\'opez-Gallifa, \inst{1}
           A. Mart\'inez-Henares, \inst{1}
           Z. T. P. Fried, \inst{4}
           B. A. McGuire, \inst{4,5}
           S. Mart\'in, \inst{6,7}
           M. A. Requena-Torres, \inst{8}
           B. Tercero, \inst{9,10}
           P. de Vicente, \inst{10}
           L. Kolesnikov\'a, \inst{11}
           E. R. Alonso, \inst{12}
           E. J. Cocinero, \inst{13,14}
           J. C. Guillemin \inst{15}
           \and
           I. Kleiner \inst{16}
          }
        \institute{Centro de Astrobiolog\'ia (CAB), INTA-CSIC, Carretera de Ajalvir km 4, Torrej\'on de Ardoz, 28850, Madrid, Spain\\
              \email{miguel.sanz.novo@cab.inta.csic.es}
        \and
        Star and Planet Formation Laboratory, Pioneering Research Institute, RIKEN, 2-1 Hirosawa, Wako, Saitama, 351-0198, Japan
        \and
        Departamento de F{\'i}sica de la Tierra y Astrof{\'i}sica, Facultad de Ciencias F{\'i}sicas, Universidad Complutense de Madrid, 28040 Madrid, Spain
        \and
        Massachusetts Institute of Technology, Cambridge, MA 02139, USA
        \and
        National Radio Astronomy Observatory, Charlottesville, VA 22903, USA
        \and
        European Southern Observatory, Alonso de C\'ordova 3107, Vitacura 763 0355, Santiago, Chile
        \and
        Joint ALMA Observatory, Alonso de C\'ordova 3107, Vitacura 763 0355, Santiago, Chile
        \and
        Department of Physics, Astronomy and Geosciences, Towson University, Towson, MD 21252, USA
        \and
        Observatorio Astronómico Nacional (OAN-IGN), Calle Alfonso XII, 3, 28014 Madrid, Spain
        \and
        Observatorio de Yebes (OY-IGN), Cerro de la Palera SN, Yebes, Guadalajara, Spain
        \and      
        Department of Analytical Chemistry, University of Chemistry and Technology, Technick\'a 5, 166 28 Prague 6, Czechia
        \and 
        Grupo de Espectroscop\'ia Molecular, Edificio Quifima, Parque Cient\'ifico UVa, Universidad de Valladolid, Valladolid, 47005 Spain.
        \and 
        Departamento de Qu\'imica F\'isica, Facultad de Ciencia y Tecnolog\'ia, Universidad del Pa\'is Vasco (EHU), 48940 Leioa, Spain.
        \and 
        Instituto Biofisika (CSIC, EHU), 48940 Leioa, Spain.
        \and Univ Rennes, Ecole Nationale Sup\'erieure de Chimie de Rennes, CNRS, ISCR-UMR6226, F-35000 Rennes,France
        \and
        Universit\'e Paris Cit\'e and Univ Paris Est Creteil, CNRS, LISA, F-75013 Paris, France
       }

\date{\today}


  \abstract
   {}
   {The tentative detection of 3-hydroxypropanal (\ch{HO(CH2)2C(O)H}) toward the Galactic center molecular cloud G+0.693-0.027 prompts a systematic survey in this source aimed at detecting all \ch{C3H6O2} isomers with available spectroscopy.}
   {We use an ultra-deep broadband spectral survey of G+0.693-0.027, carried out with the Yebes 40 m and IRAM 30 m telescopes, to conduct the astronomical search.}
   {We report the first interstellar detection of lactaldehyde (\ch{CH3CH(OH)C(O)H}) and methoxyacetaldehyde (\ch{CH3OCH2C(O)H}), together with the second detections (i.e., confirmation) of methyl acetate (\ch{CH3C(O)OCH3}) and hydroxyacetone (\ch{CH3C(O)CH2OH}), and new detections in this source of both $anti$- and $gauche$- conformers of ethyl formate (\ch{CH3CH2OC(O)H}), the latter tentatively. For these species, we derived a fractional abundance relative to H$_2$ of $\sim$(0.81, 0.24, 16, 1.6, 1.3, 1.4) $\times$ 10$^{-10}$, respectively. In contrast, neither propionic acid, \ch{CH3CH2C(O)OH}, nor glycidol, c-\ch{CH2OCHCH2OH} (i.e., the most and the least stable species within the \ch{C3H6O2} family, respectively) were detected, and we provide upper limits on their fractional abundances of $\leq$1.5 $\times$ 10$^{-10}$ and $\leq$3.7 $\times$ 10$^{-11}$. Interestingly, all \ch{C3H6O2} isomers can be synthesized through radical–radical reactions on the surface of dust grains, ultimately tracing back to CO as the parent molecule. We suggest that formation of the detected isomers is mainly driven by successive hydrogenation of CO, producing CH$_3$OH and CH$_3$CH$_2$OH as the primary parent species. Conversely, propionic acid is thought to originate from the oxygenation of CO via the HOCO intermediate, which help us rationalize its non-detection. Overall, our findings notably expand the known chemical inventory of the interstellar medium and provide direct observational evidence that increasingly complex chemistry involving O-bearing species occurs in space.}
   {}
   

\maketitle
\nolinenumbers

\section{Introduction} 
\label{sec:intro}

Structural isomerism, in which molecules share an identical molecular formula but differ in the connectivity of their constituent atoms, represents a fundamental source of chemical diversity in the interstellar medium (ISM). From a chemical point of view, this diversity translates into distinct isomers often exhibiting markedly different electronic structures, thermodynamic stabilities and spectroscopic signatures, making the identification of individual isomers a valuable tool for discriminating among their plausible formation pathways \citep{shingledecker2019,Neill:2012fr,Marcelino2021,Rivilla23,Garcia_de_la_concepcion2023,SanAndres2024,Remijan2025,Sanz-Novo2025b}.

Within this context, the \ch{C3H6O2} isomeric family has garnered particular attention in recent years from the astrochemical community. It encompasses molecules with diverse functional groups, ranging from families that are widespread in nature such as aldehydes (i.e., 2-hydroxypropanal or lactaldehyde, \ch{CH3CH(OH)C(O)H}, 3-hydroxypropanal, \ch{HO(CH2)2C(O)H}, and methoxyacetaldehyde, \ch{CH3OCH2C(O)H}), esters (i.e. methyl acetate, \ch{CH3C(O)OCH3}, and ethyl formate, \ch{CH3CH2OC(O)H}), ketones (hydroxyacetone, \ch{CH3C(O)CH2OH}), and carboxylic acids (propionic acid, \ch{CH3CH2C(O)OH}), to more structurally exotic compounds like epoxy alcohols (glycidol, c-\ch{CH2OCHCH2OH}), diols, and enediols (i.e., prop-1-ene-1,2-diol, \ch{CH3C(OH)CHOH}, prop-2-ene-1,2-diol, \ch{CH2C(OH)CH2OH}, and prop-1-ene-1,3-diol, \ch{HOCH2CHCHOH}).

Identifying which of these isomers are present in the ISM requires the synergy of accurate high-resolution laboratory rotational data, sensitive astronomical observations and astrochemical modeling. These combined efforts have thus far yielded the detection of both the \textit{anti} and \textit{gauche} conformers of ethyl formate \citep{ethyl_formate_Belloche,methyl_acetate_Orion,rivilla_chemical_2017,peng_alma_2019}, methyl acetate \citep{methyl_acetate_Orion} and hydroxyacetone \citep{hydroxyacetone_IRAS} toward various star-forming regions, as well as the recent tentative detection of 3-hydroxypropanal toward the Galactic center (GC) molecular cloud G+0.693-0.027 (hereafter G+0.693, \citealt{Fried2025}). Meanwhile, unfruitful searches have been reported for propionic acid, the global minimum in energy, and for higher-energy isomers such as methoxyacetaldehyde and lactaldehyde in Orion KL and Sagittarius B2(N) \citep{ilyushin_submillimeter_2021,Belloche2025,lactaldehyde_Alonso,methoxyacetaldehyde_Kolesnikova}.

\begin{center}
\begin{figure*}[ht]
     \centerline{\resizebox{0.975
     \hsize}{!}{\includegraphics[angle=0]{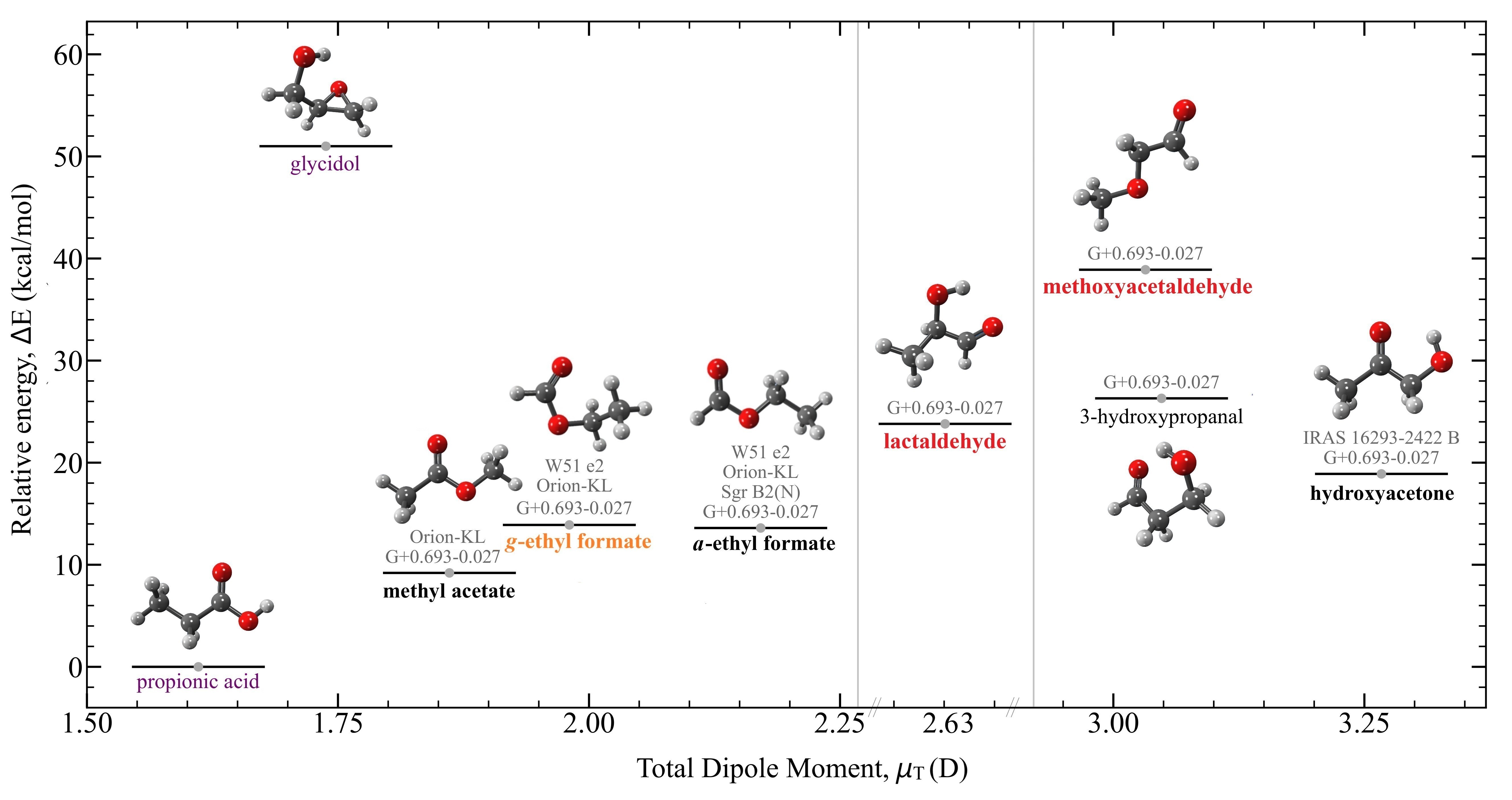}}}
     \caption{Relative energies (including zero-point energy, ZPE, corrections) plotted as a function of the total dipole moment for the targeted C$_3$H$_6$O$_2$ isomers, computed at the B2PLYPD3/aug-cc-pVTZ level of theory \citep{Grimme2011} using the Gaussian 16 program package \citep{Frisch2016}. Optimized 3D structures at the same level of theory are also shown (carbon atoms in gray, oxygen atoms in red and hydrogen atoms in white). The status of detections in G+0.693 is indicated using the following color code in the molecule name: new first detections in the ISM are shown in red, tentative detections in orange, non-detections in purple and previous detections toward different sources in black. Previous detections toward other sources of methyl acetate \citep{methyl_acetate_Orion}, \textit{anti} and \textit{gauche} ethyl formate \citep{ethyl_formate_Belloche,methyl_acetate_Orion,rivilla_chemical_2017,peng_alma_2019}, hydroxyacetone \citep{hydroxyacetone_IRAS} and 3-hydroxypropanal are also included. Also, new detections based in this work are highlighted in boldface. The 3D representations of all molecules have been visualized with IQmol ( \url{https://www.iqmol.org/}).
}
\label{f:energies}
\end{figure*}
\end{center}

In the laboratory, experiments have shown that multiple \ch{C3H6O2} isomers can form under typical astrophysical conditions. \citet{wang23} reported the production of hydroxyacetone, methyl acetate, and 3-hydroxypropanal -all of which have already been identified in the ISM- along with their enol tautomers in UV-irradiated methanol–acetaldehyde ices, offering insights into their possible formation pathways. In a subsequent study, \citet{wang24} demonstrated that the irradiation of CO–ethanol ices yields, in addition to the known interstellar species ethyl formate and 3-hydroxypropanal, other isomers still uncharted in the ISM, such as 1,3-propenediol and lactaldehyde.

In light of these findings, a systematic survey of all the spectroscopically characterized \ch{C3H6O2} isomers (see Figure \ref{f:energies}) will be essential for expanding the known interstellar chemical inventory of complex organic molecules (COMs, defined as carbon-based molecules comprised of 6 or more atoms; \citealt{Herbst2020}), and for unveiling whether a particularly complex, yet unknown underlying chemistry involving relatively large O-bearing species occurs in space. 

Over the last few years, the search for COMs has increasingly turned toward unusual environments in the GC such as the GC molecular cloud G+0.693. This cloud appears as one of the most compelling sources for the potential detection of the remaining \ch{C3H6O2} isomers, as supported by the growing inventory of COMs discovered in it (see e.g., \citealt{Rivilla23,Sanz-Novo2025a}). It includes new complex O-bearing species with one C atom such as \textit{cis}-carbonic acid (HOCOOH; \citealt{Sanz-Novo2023}), two C atoms like (\textit{Z})-1,2-ethenediol ((CHOH)$_2$, \citealt{Rivilla2022a}) and \textit{trans}-methyl formate (\ch{CH3OCHO}; \citealt{Sanz-Novo2025b}), and three C atoms such as propanol (C$_3$H$_7$OH; \citealt{Jimenez-Serra22}) and 3-hydroxypropanal \citep{Fried2025}. The latter, which was tentatively detected earlier this year, motivated us to carry out the present follow-up study.


Our analysis has confirmed the presence of six additional \ch{C3H6O2} isomers in G+0.693, including the first interstellar detection of lactaldehyde and methoxyacetaldehyde, together with the second interstellar detection (confirmation) of methyl acetate and hydroxyacetone and the fourth detection of ethyl formate in the ISM. Propionic acid and glycidol still remain undetected, but we have also reported the upper limits of their molecular column densities, setting new constraints on their abundances in the ISM.


\begin{center}
\begin{figure*}[ht]
     \centerline{\resizebox{0.8
     \hsize}{!}{\includegraphics[angle=0]{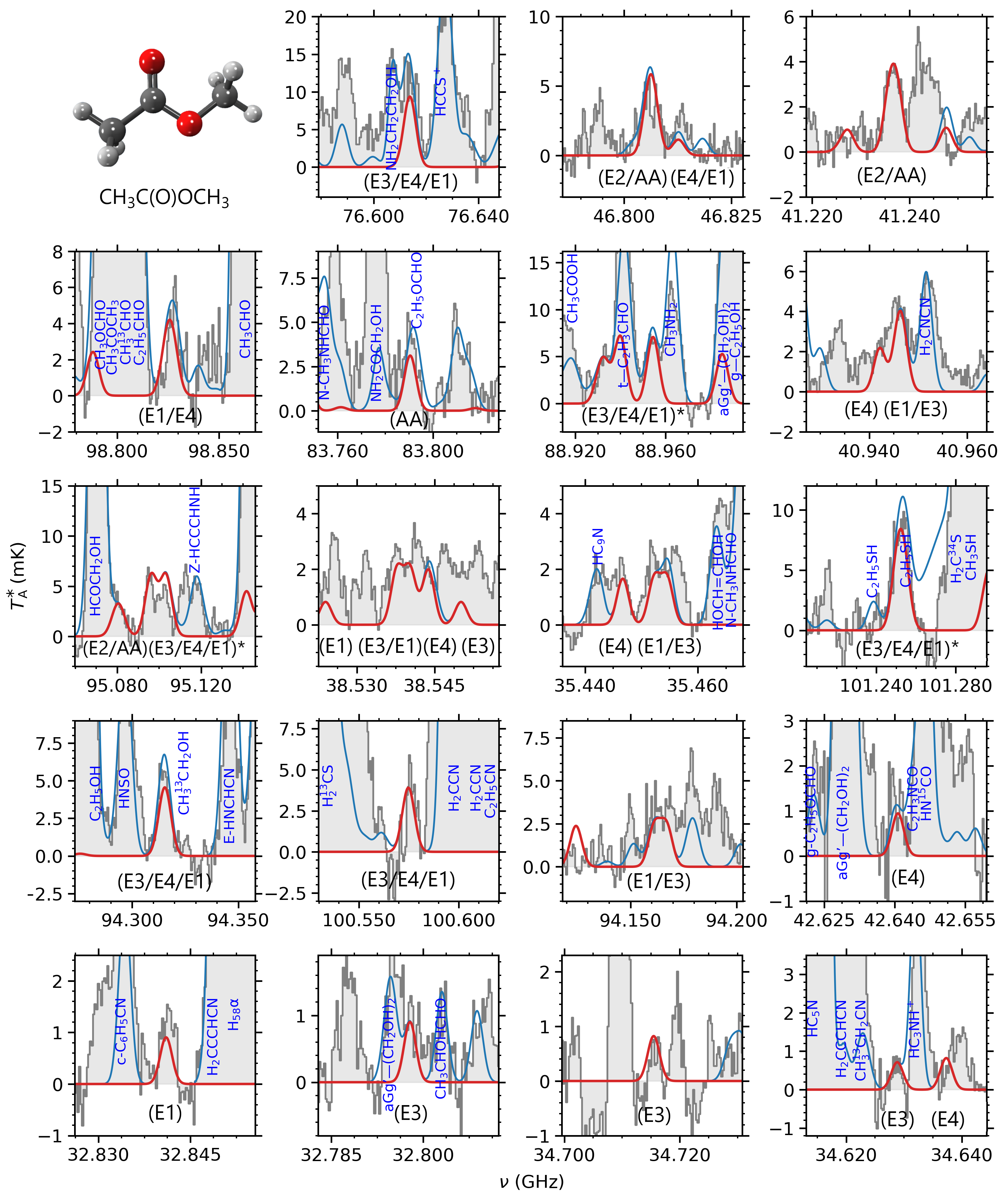}}}
     \caption{Selected unblended or slightly blended transitions of methyl acetate, CH$_{3}$C(O)OCH$_3$, identified toward G+0.693, which were used to derive the LTE physical parameters of the molecule (see text; listed in Table \ref{tab:LTEMA}). The red line shows the best LTE fit of CH$_{3}$C(O)OCH$_3$, while the blue line represents the combined emission of all molecules identified in the survey (observed spectra are shown as gray histograms). The symmetry state of the detected transitions is shown at the bottom of each panel (full list of quantum numbers listed in Table \ref{tab:LTEMA}). The * symbol indicates that there are various transitions in the same panel, each comprising the same E$_3$/E$_4$/E$_1$ symmetry states. The transitions are sorted by decreasing intensity. The molecular structure of CH$_{3}$C(O)OCH$_3$ is also shown (C atoms in gray, O atoms in red and H atoms in white). }
\label{f:LTEMA}
\end{figure*}
\end{center}


\section{Observations} 
\label{sec:obs}

The astronomical search for the \ch{C3H6O2} isomers has been performed toward the GC molecular cloud G+0.693, located approximately 55'' northeast of the well-known high-mass star-forming region Sgr B2(N). This source exhibits intermediate H$_2$ densities of $10^4$--$10^5$ $\mathrm{cm}^{-3}$ which lead to the subthermal excitation of the molecular emission, with excitation temperatures between $\sim$5--20 K (well below the kinetic temperature of the gas of 70--150 K; \citealt{requena-torres_organic_2006,zeng2018,Jimenez-Serra20,Colzi2024}). The spectra toward this source are, therefore, significantly less crowded than those observed toward hot cores, where some of the \ch{C3H6O2} isomers have been previously detected and where vibrationally excited states are also efficiently populated. This considerably reduces the line blending and facilitates the identification of new interstellar species through a smaller number of rotational transitions.

In this study, we employed an ultra-deep spectral survey of this source that combines centimeter- and millimeter-wavelength observations acquired with two facilities. The $Q$-band range (31.075–50.424 GHz) was surveyed using the Yebes 40-m telescope in Guadalajara, Spain (projects 20A008 and 21A014), while the IRAM 30-m telescope in Granada, Spain, provided complementary coverage in two broadband millimeter-wavelength frequency windows (71.8–116.7 GHz, 124.8–175.5 GHz, see \citealt{Rivilla23,Sanz-Novo2023} and references therein; projects 172-18, 018-19, 133-19, 123-22). Data were collected in position-switching mode, centered at $\alpha$ = 17$^{\rm h}$47$^{\rm m}$22$^{\rm s}$, $\delta$ = $-28^{\circ}21^{\prime}27^{\prime\prime}$ (J2000), with an off-source reference located at $\Delta\alpha = -885^{\prime\prime}$ and $\Delta\delta = 290^{\prime\prime}$. The Yebes telescope’s half-power beam width (HPBW) spans $\sim$55$^{\prime\prime}$–35$^{\prime\prime}$ from 31 to 50 GHz \citep{tercero2021}, whereas the IRAM 30-m HPBW ranges from $\sim$32$^{\prime\prime}$ to $\sim$10$^{\prime\prime}$ across the targeted frequencies. Given that molecular emission toward G+0.693 is extended compared to the beam size of the telescope \citep{Jones2012,Li2020,Zheng2024}, we report all spectra in antenna temperature units ($T_A^*$). Overall, we have achieved root mean square (rms) noise levels lying between 0.25$-$0.9 mK across the whole $Q$-band,
0.5$-$2.5 mK at 3 mm, and 1.0$-$1.6 mK at 2 mm. Further details on these observations and their reduction are available in \citet{Rivilla23} and \citet{Sanz-Novo2023}. 


\begin{table*}
\tabcolsep 1.0pt
\centering
\caption{Derived physical parameters for the \ch{C3H6O2} isomers that we have searched for toward G+0.693-0.027.}
\begin{tabular}{ c c c c c c c c c c}
\hline
\hline
 Molecule & Formula & $\Delta E$$^{a}$ & $\mu$(tot)$^a$ & $N$  &  $T_{\rm ex}$ & $v$$_{\rm LSR}$ & FWHM  & Abundance$^b$ &  Status$^c$  \\
          &         & (kcal mol$^{-1}$/K) & (D) &     ($\times$10$^{13}$ cm$^{-2}$) & (K) & (km s$^{-1}$) & (km s$^{-1}$) & ($\times$10$^{-10}$)  &  \\
\hline
propionic acid  & \ch{CH3CH2C(O)OH} & 0.0 / 0.0 & 1.6 & $\leq$ 2 & 15$^d$ & 68$^d$ & 21$^d$ & $\leq$ 1.5 & $nd$\\
methyl acetate  & \ch{CH3C(O)OCH3}  & 9.2 / 4646 & 1.9 & 22 $\pm$ 2 &  15 $\pm$ 2 & 68 $\pm$ 1  & 21$^d$  & 16 $\pm$ 3 & $d$ \\
\textit{anti}-ethyl formate &  $a$-\ch{CH3CH2OC(O)H} & 13.6 / 6820 & 2.2 & 1.82 $\pm$ 0.15 & 15$^d$  &  68$^d$  & 21$^d$  & 1.3 $\pm$ 0.2 & $d$ \\
\textit{gauche}-ethyl formate & $g$-\ch{CH3CH2OC(O)H}  & 13.9 / 7014 & 2.0 & 1.9 $\pm$ 0.2  & 15$^d$ & 68$^d$ & 21$^d$ &  1.4 $\pm$ 0.3  & $td$\\
hydroxyacetone & \ch{CH3C(O)CH2OH}  & 18.9 / 9510 & 3.3 & 2.11 $\pm$ 0.13 & 15$^d$ & 68$^d$ & 21$^d$ & 1.6 $\pm$ 0.2  & $d$\\
lactaldehyde & \ch{CH3CH(OH)C(O)H} & 23.8 / 11997 & 2.6 &  1.10 $\pm$ 0.13 & 15$^d$ & 68$^d$ & 21$^d$ & 0.81 $\pm$ 0.16 & $fd$ \\
3-hydroxypropanal  & \ch{HO(CH2)2C(O)H} & 26.3 / 13239 & 3.0 & 0.86 $\pm$ 0.14 & 12$^{d}$ & 69$^{d}$ & 21$^{d}$ & 0.64 $\pm$ 0.14 & $td$\\
methoxyacetaldehyde  & \ch{CH3OCH2C(O)H} & 38.9 / 19581& 3.0 & 0.33 $\pm$ 0.03 & 15$^d$ & 68$^d$ & 21$^d$ & 0.24 $\pm$ 0.04 & $fd$ \\
glycidol  & c-\ch{CH2OCHCH2OH} & 51.0 / 25674 &  1.7 & $\leq$0.5 & 15$^d$ & 68$^d$ & 21$^d$ & $\leq$0.37 & $nd$ \\
\hline
\end{tabular}
\label{tab:comparison}
\vspace{0mm}
\vspace*{-0.5ex}
\tablefoot{$^a$ Relative energies (including ZPE corrections) and total electric dipole moments are computed at the B2PLYPD3/aug-cc-pVTZ level of theory \citep{Grimme2011} using the Gaussian 16 program package \citep{Frisch2016}. $^b$ We adopted $N_{\rm H_2}$ = 1.35$\times$10$^{23}$ cm$^{-2}$, from \citet{martin_tracing_2008}, assuming an uncertainty of 15\% of its value. $^c$ Status of the search for each species toward G+0.693: $d$ = detection, $fd$ = first detection, $td$ = tentative detection, $nd$ = non-detection. $^d$ Values fixed in the fit.}
\label{tab:g0693}
\end{table*}


\section{Results} 
\label{sec:detection}

\subsection{LTE analysis}
\label{subsec:LTEanalysis}

We carried out a local thermodynamic equilibrium (LTE) analysis of all \ch{C3H6O2} isomers for which rotational spectroscopic data are available, using the Spectral Line Identification and Modeling (SLIM) tool (version from 15 June 2024) within the \textsc{Madcuba} software package \citep{martin2019}. This tool allows us to produce a synthetic spectrum of each target species and compare it directly with the observed spectra, as well as to evaluate the emission of all the molecules previously identified in the current molecular line survey (i.e., over 140 species).

To obtain the best-fit LTE model for each molecule, we used the \textsc{Autofit} tool within SLIM \citep{martin2019} and performed a nonlinear least-squares LTE fit of a selected subset of clean transitions to the observed data using the Levenberg-Marquardt algorithm. We note that heavily blended lines, contaminated by the emission of previously identified species, were excluded from the analysis. For the fit, we can leave as free parameters: the molecular column density ($N$), the excitation temperature ($T_{\rm ex}$), the radial velocity ($v$$_{\rm LSR}$) and the linewidth (FWHM). As described in the following subsections, we fixed certain parameters for some of the studied species to that derived for the most abundant \ch{C3H6O2} isomer, methyl acetate, which also presents the brightest transitions, to ensure convergence of the \textsc{Autofit}.

In addition, to derive the fractional abundances with respect to molecular hydrogen for all targeted molecules, we adopted $N$(H$_2$) = 1.35$\times$10$^{23}$ cm$^{-2}$ from \citet{martin_tracing_2008}, assuming an uncertainty of 15\% of its value. For comparison and completeness, we list in Table \ref{tab:comparison} the physical parameters derived for all \ch{C3H6O2} isomers, including those reported for 3-hydroxypropanal in \cite{Fried2025}.


\subsection{Detection of Methyl acetate}

Methyl acetate, CH$_{3}$C(O)OCH$_3$, is the second most stable species within the \ch{C3H6O2} isomeric family, and was first detected in Orion KL by \cite{methyl_acetate_Orion}. However, this molecule has not yet been identified in any other astronomical environment. Here, we searched for CH$_{3}$C(O)OCH$_3$ toward G+0.693, which resulted in an unequivocal detection (see Figure \ref{f:LTEMA}), thus pointing to a broader presence of methyl acetate in the ISM. For the astronomical search, we employed the spectroscopic catalog derived from the original analysis using the BELGI-$C_{\rm s}$-2Tops code \citep{methyl_acetate_Orion}, which we converted into the common SPCAT/SPFIT format \citep{Pickett1991} (see Appendix Sect. \ref{rot_backgr}). This molecule exhibits two non-equivalent CH$_3$ internal rotors, which split the rotational energy levels into five substates (i.e., AA, AE, EA, and two EE species, E$_3$ and E$_4$, with statistical weights of 16, 16, 16, 8 and 8, respectively). In the transformed catalog, these symmetry species are labeled using the so-called vibrational quantum number as follows: 0 = AA, 1 = E$_1$ (EA), 2 = E$_2$ (AE), 3 = E$_3$ (EE), 4 = E$_4$ (EE) (see Table \ref{tab:LTEMA}).

In Figure \ref{f:LTEMA}, we show a selection of the brightest and least contaminated lines of CH$_{3}$C(O)OCH$_3$ that have been detected toward G+0.693, which were subsequently used to derive the physical parameters of the targeted molecular emission. Moreover, the fine structure corresponding to transitions belonging to the five symmetry states is, in some cases, partially resolved, giving rise to easily recognizable patterns, while in other cases it is fully merged into a single, broader spectral feature due to the typically broad linewidth of the molecular line emission measured toward G+0.693 (i.e., FWHM $\sim$ 15$-$20 km s$^{-1}$; \citealt{requena-torres_organic_2006,requena-torres_largest_2008,zeng2018,Rivilla2022c}). However, this “autoblending" likewise facilitates the detection because it strengthens the intensity of the resulting line cluster (e.g., 16$_{1,16}$--15$_{1,15}$ transition at $\sim$101.2512 GHz). 

To yield the best LTE modeling for CH$_{3}$C(O)OCH$_3$, we used the \textsc{Autofit} tool for the clean subset of transitions shown in Figure \ref{f:LTEMA} (spectroscopic information reported in Table \ref{tab:LTEMA}). For the fit, we fixed the FWHM to a value of 21 km s$^{-1}$, which reproduces well the brightest and cleanest transitions, while the $N$, $T_{\rm ex}$ and $v$$_{\rm LSR}$ were left as free parameters. We obtained a $N$ = (2.2 $\pm$ 0.2) $\times$ 10$^{14}$ cm$^{-2}$, which corresponds to a fractional abundance with respect to molecular hydrogen of (1.6$\pm$ 0.3) $\times$ 10$^{-9}$. Furthermore, we derived a $T_{\rm ex}$ = 15 $\pm$ 2 K and a $v$$_{\rm LSR}$ = 68 $\pm$ 1 km s$^{-1}$, consistent with those employed for the analysis of its isomer 3-hydroxypropanal \citep{Fried2025} (see Table \ref{tab:comparison}). We note that a change of 1-2 K in the $T_{\rm ex}$ only slightly affects the derived $N$, and in all cases this variation is smaller than the uncertainty assumed in the derivation of the fractional abundance with respect to H$_2$ (i.e., 15\%).

\begin{center}
\begin{figure*}[ht]
     \centerline{\resizebox{0.8
     \hsize}{!}{\includegraphics[angle=0]{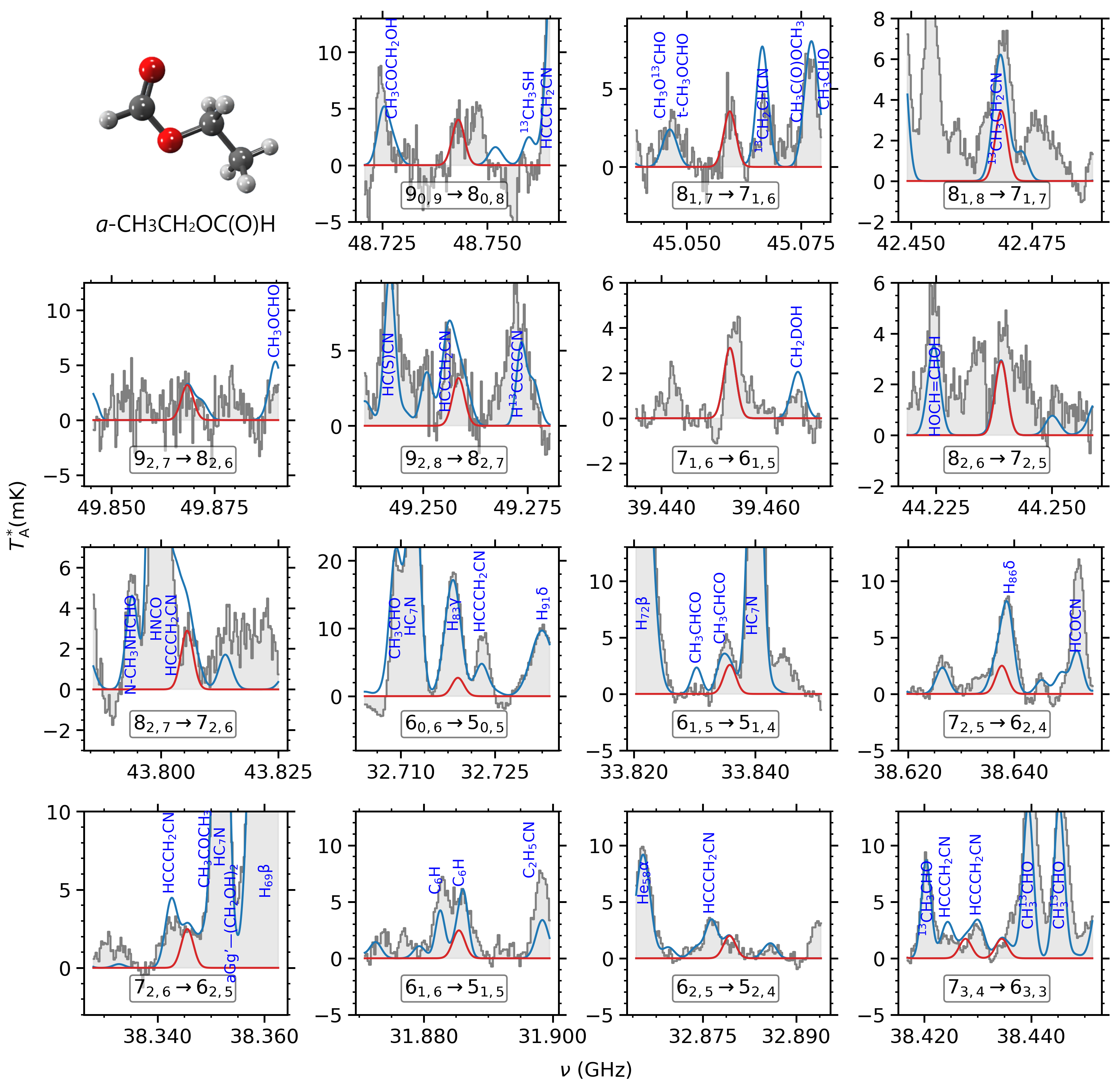}}}
     \caption{Transitions of \textit{a}-ethyl formate, $a$-\ch{CH3CH2OC(O)H}, identified toward G+0.693, which were used to derive the LTE physical parameters of the molecule (see text; listed in Table \ref{tab:aethylformate}). The red line shows the best LTE fit of $a$-\ch{CH3CH2OC(O)H}, while the blue line represents the combined emission of all molecules identified in the survey, including  $a$-\ch{CH3CH2OC(O)H} (observed spectra are shown as gray histograms). The quantum numbers for each transition are shown at the bottom of each panel. The transitions are sorted by decreasing intensity. The molecular structure of $a$-\ch{CH3CH2OC(O)H} is also shown (C atoms in gray, O atoms in red and H atoms in white).}
\label{f:LTEaEF}
\end{figure*}
\end{center}


\subsection{Detection of anti-Ethyl formate and tentative detection of gauche-Ethyl formate}

Following in energy, we find ethyl formate (\ch{CH3CH2OC(O)H}). Its two lowest-in-energy conformers \footnote{Structures that share the same molecular formula and connectivity of atoms, but differ in their 3D-arrangement in space}, $anti$ and $gauche$, are separated by 65 $\pm$ 21 cm$^{-1}$ (or 94 K; \citealt{Riveros:1967ic}) with the $anti$ configuration being the global minimum in energy. The \textit{anti} conformer (hereafter $a$-\ch{CH3CH2OC(O)H}) has been detected so far in three hot molecular cores: Sgr B2(N), Orion KL and W51 e2 \citep{ethyl_formate_Belloche,methyl_acetate_Orion,rivilla_chemical_2017,peng_alma_2019}, with the latter two sources also harboring \textit{gauche} \ch{CH3CH2OC(O)H} (hereafter $g$-\ch{CH3CH2OC(O)H}) in an approximately 1:1 abundance ratio.

In this work, we report a new detection of $a$-\ch{CH3CH2OC(O)H} toward G+0.693, while $g$-\ch{CH3CH2OC(O)H} is tentatively detected due to non-negligible line blending for the majority of lines (see Figures \ref{f:LTEaEF} and \ref{f:LTEgEF}). To perform the corresponding LTE analysis to the $a$-\ch{CH3CH2OC(O)H} emission, we used the spectroscopic catalog from the Cologne Database for Molecular Spectroscopy (CDMS, entry 074514; \citealt{endres_cologne_2016}), which is based on the laboratory data by \citet{Riveros:1967ic,Demaison:1984if,Medvedev:2009fs}. For the $g$-conformer, we employed a new spectroscopic entry using the aforementioned data, since the CDMS entry includes the rotational lines of $g$-\ch{CH3CH2OC(O)H} shifted by the corresponding $\Delta$E($a$-$g$), thus not being suitable for the analysis in regions whose molecular emission is well described by low excitation temperatures. Hence, we considered each conformer as distinct molecular species, as previously done for other molecules in G+0.693 (see e.g., \citealt{Jimenez-Serra22,Sanz-Novo2023,Sanz-Novo2025b}).

After inspecting the astronomical data, various unblended or slightly blended $a$-type transitions ranging from $J$ = 6 to $J$ = 9, with $K$$_a$ $<$ 3 and belonging to $a$-\ch{CH3CH2OC(O)H} were detected. A selection of the brightest and cleanest transitions, which were subsequently used in the LTE fit with \textsc{Madcuba}, are shown in Figure \ref{f:LTEaEF} (spectroscopic information reported in Table \ref{tab:aethylformate}). For completeness, we also show other transitions that appear blended, but their predicted intensities are nevertheless required to reproduce the observed spectra once the contribution of all species previously identified toward G+0.693 is taken into account. In this case, we fixed the $T_{\rm ex}$, $v$$_{\rm LSR}$ and FWHM to the values derived for methyl acetate (i.e., $T_{\rm ex}$ = 15 K, $v$$_{\rm LSR}$ = 68 km s$^{-1}$, FWHM = 21 km s$^{-1}$), leaving the $N$ as the only free parameter in the \textsc{Autofit}. We derived a $N$ = (1.82 $\pm$ 0.15) $\times$ 10$^{13}$ cm$^{-2}$, which yields a fractional abundance with respect to H$_2$ of (1.3$\pm$ 0.2) $\times$ 10$^{-10}$.

Meanwhile, for $g$-\ch{CH3CH2OC(O)H} we detected two unblended lines along with seven additional lines showing slight blending. Therefore, the overall detection of this conformer remains tentative (see Figure \ref{f:LTEgEF}, spectroscopic information listed in Table \ref{tab:gethylformate}). Following the same fitting approach as for the $anti$ conformer, we obtained a $N$ = (1.9 $\pm$ 0.2) $\times$ 10$^{13}$ cm$^{-2}$, which corresponds to a fractional abundance with respect to H$_2$ of (1.4 $\pm$ 0.3) $\times$ 10$^{-10}$, and does not produce any overly bright feature for other transitions. Therefore, based on the derived abundance, we infer an $anti$/$gauche$ ratio of 0.9$\pm$0.1, in excellent agreement with the ratio reported previously toward Orion KL and W51 e2 (of $\sim$1; \citealt{methyl_acetate_Orion,rivilla_chemical_2017}). Furthermore, the total abundance of \ch{CH3CH2OC(O)H} (i.e., the sum of the abundance of both $a$- and $g$-conformers) is a factor of 5.9$\pm$0.6 lower than that of CH$_{3}$C(O)OCH$_3$ in G+0.693, yielding a CH$_{3}$C(O)OCH$_3$/\ch{CH3CH2OC(O)H} abundance ratio, comparable to that observed previously toward Orion KL (of $\sim$5; \citealt{methyl_acetate_Orion}).


\subsection{Detection of Hydroxyacetone}

\begin{center}
\begin{figure*}[ht]
     \centerline{\resizebox{0.8
     \hsize}{!}{\includegraphics[angle=0]{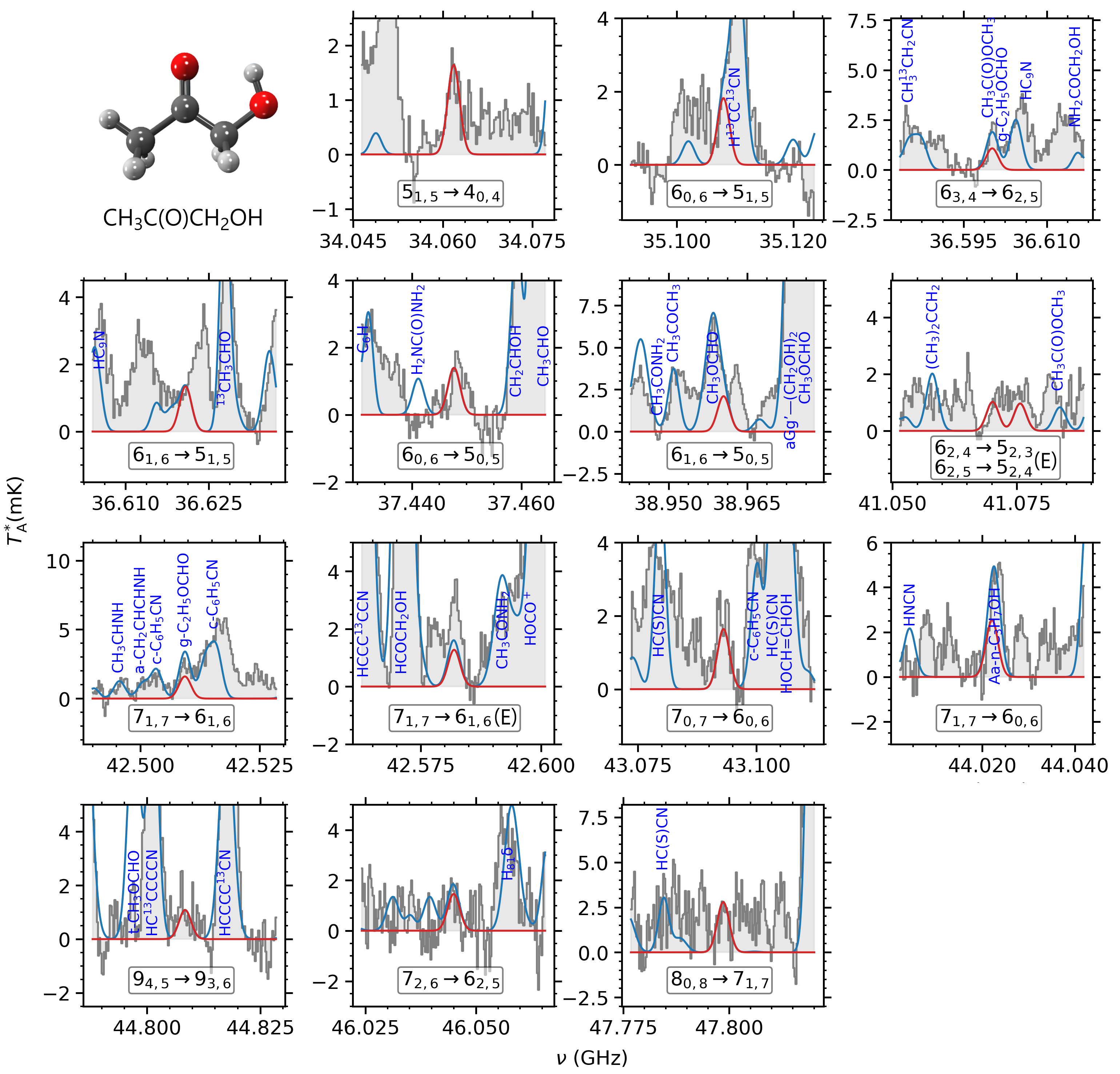}}}
     \caption{Transitions of hydroxyacetone, CH$_{3}$C(O)CH$_2$OH, detected toward G+0.693 (see text; listed in Table \ref{tab:hydroxyacetone}). The red line shows the best LTE fit of CH$_{3}$C(O)CH$_2$OH, while the blue line represents the combined emission of all molecules identified in the survey, including CH$_{3}$C(O)CH$_2$OH (observed spectra are shown as gray histograms). The quantum numbers for each transition are shown at the bottom of each panel (A state lines unless stated otherwise). The transitions are sorted by increasing frequency. The molecular structure of CH$_{3}$C(O)CH$_2$OH is also shown (C atoms in gray, O atoms in red and H atoms in white).}
\label{f:LTEHA}
\end{figure*}
\end{center}

To date, hydroxyacetone (CH$_{3}$C(O)CH$_2$OH; the fourth most stable structural isomer) has only been detected toward the protostar IRAS 16293–2422B \citep{hydroxyacetone_IRAS}. We have conducted a new search for the molecule toward G+0.693, yielding the second interstellar detection (first one outside a hot core), and confirming its presence in the ISM.

To carry out its search, we used the spectroscopic entry of the ground vibrational state of CH$_{3}$C(O)CH$_2$OH included in the Jet Propulsion Laboratory (JPL) database (entry 74003; \citealt{pickett1998}), based on high-resolution microwave, millimeter- and submillimeter-wave laboratory data by \cite{Braakman2010}. The molecule presents a low barrier CH$_3$ internal rotor, which causes the splitting of the energy levels into the A and E symmetry states. In this case, we detected several nearly complete progressions of $R$-branch $a$-type transitions ranging from $J$ = 5 to $J$ = 8 (a total of six fully unblended lines), with $K$$_a$ = 0,1 and belonging to CH$_{3}$C(O)CH$_2$OH, along with few additional higher $K$$_a$ lines (see Figure \ref{f:LTEHA}, spectroscopic information reported in Table \ref{tab:hydroxyacetone}). To conduct the LTE analysis, we again left the $N$ as the only free parameter and fixed the rest of parameters to those obtained for methyl acetate (listed in Table \ref{tab:comparison}). We retrieved a $N$ = (2.11 $\pm$ 0.13) $\times$ 10$^{13}$ cm$^{-2}$, which translates into a fractional abundance with respect to H$_2$ of (1.6$\pm$ 0.2) $\times$ 10$^{-10}$. Based on the derived abundance, CH$_{3}$C(O)CH$_2$OH is a factor of 10$\pm$1 less abundant than CH$_{3}$C(O)OCH$_3$, and 1.8$\pm$0.1 times less abundant than \ch{CH3CH2OC(O)H}.


\subsection{Detection of Lactaldehyde}
\label{subsec:LA}

\begin{center}
\begin{figure*}[ht]
     \centerline{\resizebox{0.8
     \hsize}{!}{\includegraphics[angle=0]{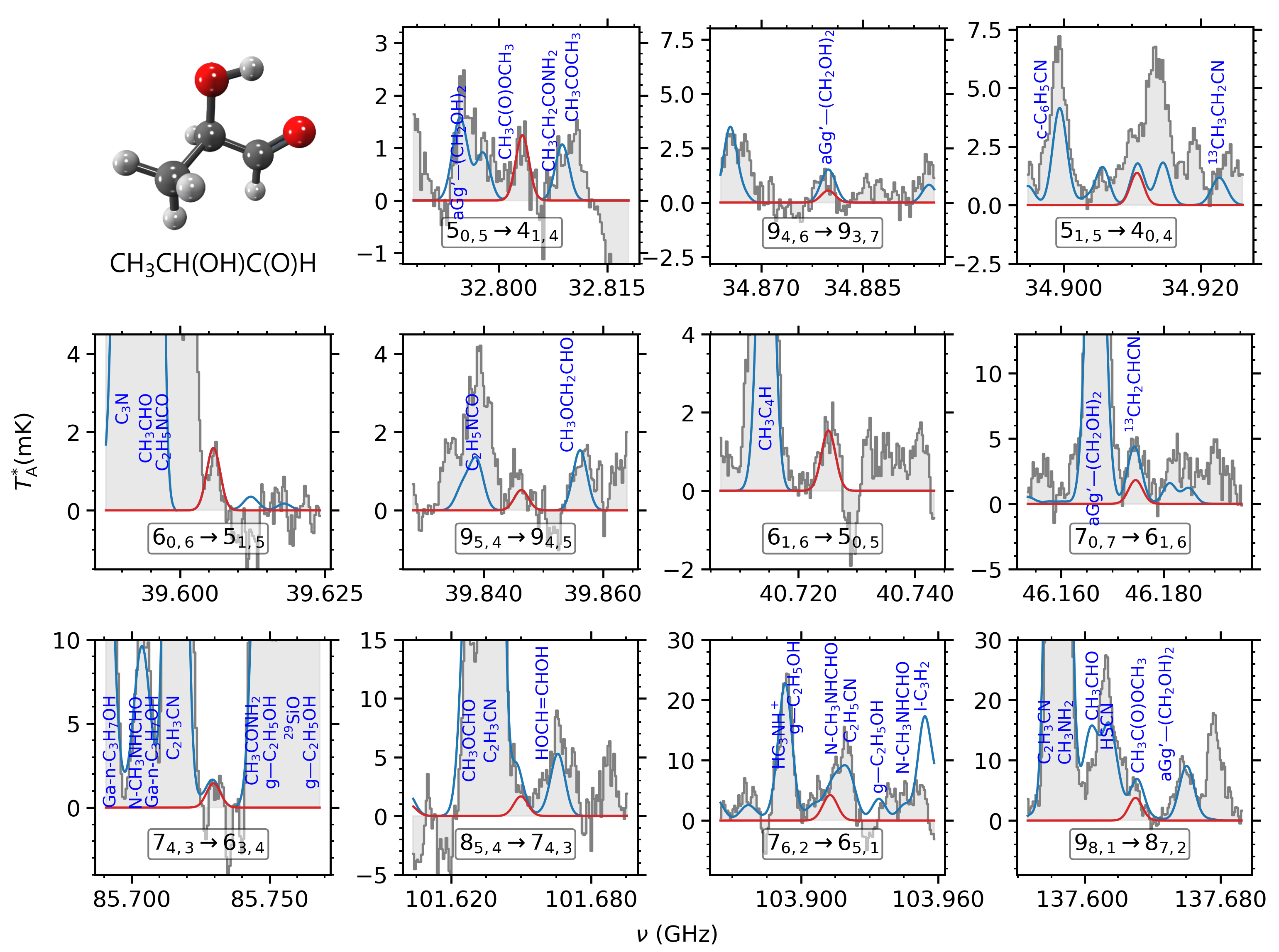}}}
     \caption{Transitions of lactaldehyde, \ch{CH3CH(OH)C(O)H}, identified toward G+0.693 (see text; listed in Table \ref{tab:lact}). The red line shows the best LTE fit of \ch{CH3CH(OH)C(O)H}, while the blue line represents the combined emission of all molecules identified in the survey, including \ch{CH3CH(OH)C(O)H} (observed spectra are shown as gray histograms). The quantum numbers for each transition are shown at the bottom of each panel. For panels with multiple lines, we show the quantum number of the lowest-frequency transition. The transitions are sorted by increasing frequency. The molecular structure of \ch{CH3CH(OH)C(O)H} is also shown (C atoms in gray, O atoms in red and H atoms in white).}
\label{f:LTELA}
\end{figure*}
\end{center}

For lactaldehyde (2-hydroxypropanal, \ch{CH3CH(OH)C(O)H}; ranked as the fifth most stable structural isomer), we incorporated the high-resolution rotational data collected by \cite{lactaldehyde_Alonso} into \textsc{Madcuba}, which corresponds to entry 074519 from CDMS \citep{endres_cologne_2016}. The exquisite sub-mK sensitivity of the present spectral survey toward G+0.693 allowed us to detect three clear and clean emission features that could be unambiguously ascribed to the 5$_{0,5}$ -- 4$_{1,4}$, 6$_{0,6}$ -- 5$_{1,5}$ and 6$_{1,6}$ -- 5$_{0,5}$ transitions of \ch{CH3CH(OH)C(O)H} (see Figure \ref{f:LTELA}, spectroscopic information reported in Table \ref{tab:lact}). These correspond to the brightest $K$$_a$ = 0,1 $R$-branch $b$-type transitions that fall within the ultra-deep $Q$-band data, all of which appear free of blending and exhibit good integrated signal-to-noise (S/N) ratios (S/N = 12, 16 and 18, respectively; rms = 0.37 mK, 0.34 mK, and 0.43 mK) \footnote{The signal to noise ratio is calculated from the integrated signal ($\int$ $T$$\mathrm{_A^*}$d$v$) and noise level $\sigma$ = rms $\times$ $\sqrt{\delta v \times \mathrm{FWHM}}$, where $\delta$$v$ is the velocity resolution of the spectra and the FWHM is fitted from the data (i.e., 21 km s$^{-1}$).}. We also detected the adjacent 7$_{0,7}$ -- 6$_{1,6}$ transition, as well as few additional higher $K$$_a$ lines that appear blended but reproduce the observed spectra well once the contribution from all other identified species is considered (blue solid line). We note that \ch{CH3CH(OH)C(O)H} is a near-prolate asymmetric top, exhibiting a dominant $b$-type spectrum ($\mu$$_a$ = 1.1 D, $\mu$$_b$ = 2.2 D, $\mu$$_c$ = 0.93; computed at the B2PLYPD3-aug-cc-pVTZ level). The fact that A $\gg$ B, C implies that the strongest $R$-branch $b$-type progressions (e.g., $J'_{1,J'} \rightarrow J''_{0,J''}$, which appear at frequencies close to $A + C + 2CJ''$) are shifted toward higher frequencies compared to the $a$-type $R$-branch lines detected for other isomers, which occur near $(B + C)(J'' + 1)$. Consequently, fewer intense transitions of \ch{CH3CH(OH)C(O)H} are expected within the $Q$-band, where the sensitivity is the highest (see \citealt{Rivilla23,Sanz-Novo2023,Sanz-Novo2025a}). Accordingly, only a few unblended or slightly blended transitions are detected at higher frequencies (e.g., 3 mm window; see last row of Figure \ref{f:LTELA}), and the rest appear either heavily blended or the derived S/N is significantly lower. Nevertheless, in all cases the LTE model is consistent with the observed spectra, and we note that no missing lines of \ch{CH3CH(OH)C(O)H} are detected across the entire survey.

Following the same fitting approach as described above for other isomers, the \textsc{Autofit} analysis yielded a $N$ = (1.10 $\pm$ 0.13) $\times$ 10$^{13}$ cm$^{-2}$, which translates into an abundance of (8.1 $\pm$ 1.6) $\times$10$^{-11}$. Accordingly, we find that lactaldehyde is 20$\pm$3, 3.4$\pm$0.4, and 1.9$\pm$0.2 times less abundant than methyl acetate, ethyl formate and hydroxyacetone, respectively, and is only slightly more abundant (a factor of 1.3$\pm$0.2) than its positional isomer 3-hydroxypropanal in G+0.693 \citep{Fried2025}.


\subsection{Detection of Methoxyacetaldehyde}

\begin{center}
\begin{figure*}[ht]
     \centerline{\resizebox{0.8
     \hsize}{!}{\includegraphics[angle=0]{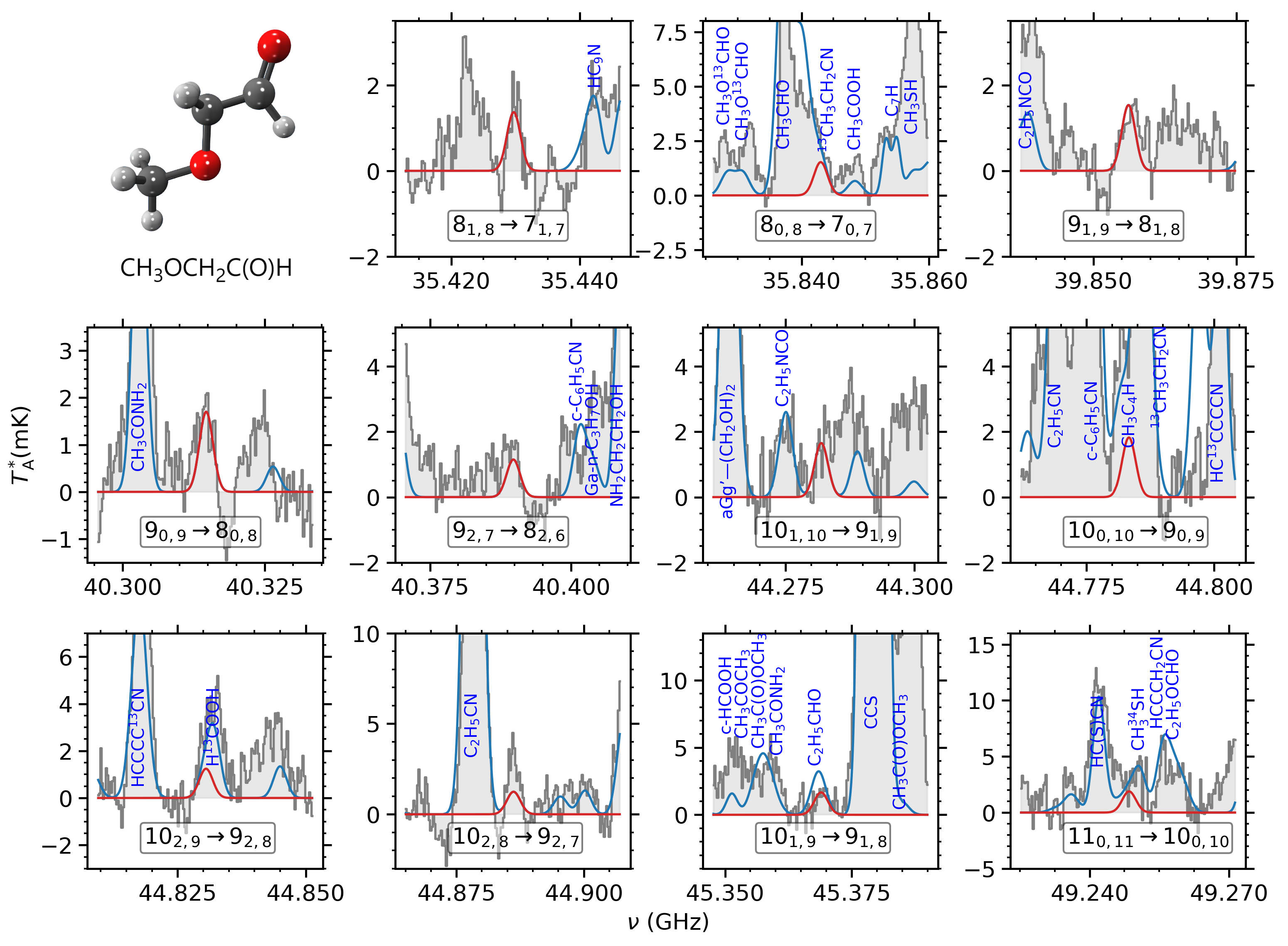}}}
      \caption{Transitions of methoxyacetaldehyde, \ch{CH3OCH2C(O)H}, detected toward G+0.693 (collected in Table \ref{tab:meta}). The red line shows the best LTE fit of \ch{CH3OCH2C(O)H}, while the blue line represents the combined emission of all molecules identified in the survey, including \ch{CH3OCH2C(O)H} (observed spectra are shown as gray histograms). The quantum numbers for each transition are shown at the bottom of each panel. The transitions are sorted by increasing frequency. The molecular structure of \ch{CH3OCH2C(O)H} is also shown (C atoms in gray, O atoms in red and H atoms in white).}
\label{f:LTEMeta}
\end{figure*}
\end{center}

Methoxyacetaldehyde (\ch{CH3OCH2C(O)H}) has been previously searched for, without success, toward Orion KL and Sgr B2, as well as toward the dark cloud Barnard 1 \citep{methoxyacetaldehyde_Kolesnikova}. To perform the astronomical search, we implemented the spectroscopic entry into \textsc{Madcuba} (CDMS entry 074517 \citealt{endres_cologne_2016}, based on \citealt{methoxyacetaldehyde_Kolesnikova}). Although \ch{CH3OCH2C(O)H} also presents methyl internal rotation, its $V$$_3$ barrier is higher than that of methyl acetate and the A/E splittings were not resolved at 3mm with a conventional absorption spectrometer \citep{methoxyacetaldehyde_Kolesnikova} or with the resolution of the present survey \citep{Rivilla23,Sanz-Novo2023}.

An inspection of the data of G+0.693 revealed a clear pattern of $R$-branch $a$-type transitions belonging to \ch{CH3OCH2C(O)H} (see Figure \ref{f:LTEMeta}). Among these, the 8$_{1,8}$ -- 7$_{1,7}$, 9$_{1,9}$ -- 8$_{1,8}$ and 9$_{0,9}$ -- 8$_{0,8}$ transitions are fully unblended and show a good integrated S/N ratio (S/N = 12, 14 and 14, respectively; rms = 0.44 mK, 0.48 mK, and 0.48 mK). We also identified the 8$_{0,8}$ -- 7$_{0,7}$, 9$_{2,7}$ -- 8$_{2,6}$ and 10$_{1,10}$ -- 9$_{1,9}$ transitions, which are slightly contaminated with $^{13}$CH$_2$CHCN and two U-lines, respectively, as well as few additional $K$$_a$ = 0, 1 and 2 transitions that appear more blended. Nevertheless, all these transitions help to reproduce the observations once the emission from all other molecules previously identified toward G+0.693 are taken into account, further reinforcing the detection. The spectroscopic information of all these transitions is reported in Table \ref{tab:meta}. We then used the aforementioned unblended and slightly blended transitions of \ch{CH3OCH2C(O)H} to carry out the LTE analysis. We retrieved a $N$ = (3.3 $\pm$ 0.3) $\times$ 10$^{12}$ cm$^{-2}$ with the \textsc{Autofit} tool. This corresponds to a fractional abundance with respect to H$_2$ of (2.4 $\pm$ 0.4) $\times$10$^{-11}$. Consequently, \ch{CH3OCH2C(O)H} shows the lowest abundance among all C$_3$H$_6$O$_2$ isomers, being a factor of 67$\pm$9, 11$\pm$1, 6.4$\pm$0.7, 3.3$\pm$0.4 and 2.6$\pm$0.5 less abundant than methyl acetate, ethyl formate, hydroxyacetone, lactaldehyde and 3-hydroxypropanal, respectively, in G+0.693.


\subsection{Non-detection of Propionic acid and Glycidol}

Propionic acid, \ch{CH3CH2C(O)OH}, besides being the most thermodynamically favored  C$_3$H$_6$O$_2$ isomer, also exhibits the lowest total dipole moment within this molecular family (i.e., $\mu$$_a$ = 0.2 D, $\mu$$_b$ = 1.6 D; computed at the B2PLYPD3-aug-cc-pVTZ level). It has been previously searched for toward Orion KL and Sgr B2(N) \citep{ilyushin_submillimeter_2021}, however, the molecule remains undetected. Here, we searched for \ch{CH3CH2C(O)OH} toward G+0.693 using a catalog based on the spectroscopy reported in \cite{Jaman2015}, but did not achieve a detection. Although many transitions still leave room for the possible presence of the molecule, the majority of the brightest transitions are affected by significant line blending (see Figure \ref{f:nondetectionPA}). To derive the upper limits for its column density, we fixed the $T_{\rm ex}$, $v$$_{\rm LSR}$ and FWHM to that of methyl acetate (listed in Table \ref{tab:comparison}) and increased $N$ until the LTE model (including also the contribution from other species) visually matched the observed spectra for the brightest $K$$_a$ = 1 transition, 8$_{1,8}$ -- 7$_{0,7}$ (at 49.3424 GHz). We estimated a $N$ $\leq$ 2 $\times$ 10$^{13}$ cm$^{-2}$, which yields a fractional abundance with respect to H$_2$ of $\leq$ 1.5 $\times$ 10$^{-10}$.

We also targeted the cyclic species glycidol (c-\ch{CH2OCHCH2OH}), which, to our knowledge, has not been searched for in the ISM to date. We used a newly generated catalog in SPFIT/SPCAT format, based on the spectroscopic parameters reported by \citet{Demaison2012}, and implemented it into \textsc{Madcuba}. After analyzing the data, we did not detect any bright, unblended transition of c-\ch{CH2OCHCH2OH} (see Figure \ref{f:nondetectionGly}). We note that this species presents a relatively low total dipole moment (i.e., the second lowest after \ch{CH3CH2C(O)OH}; $\mu$$_a$ = 0.7 D, $\mu$$_b$ = 1.3 D, $\mu$$_c$ = 0.8 D, computed at the B2PLYPD3-aug-cc-pVTZ level), which will hamper its detection. To derive the line-integrated 3$\sigma$ upper limit to its column density ($\sigma$ being the rms noise of the spectra), we used the 5$_{5,1}$ -- 4$_{4,0}$ and 5$_{5,0}$ -- 4$_{4,1}$ transitions (fully coalesced and located at $\sim$97.0723 GHz), which is the brightest feature predicted for c-\ch{CH2OCHCH2OH}, only negligibly blended with \ch{CH3(O)OCH3}. We obtained a $N$ $\leq$ 0.5 $\times$ 10$^{13}$ cm$^{-2}$, which translates into an abundance upper limit of $\leq$3.7$\times$10$^{-11}$.


\begin{center}
\begin{figure*}[ht]
    \centering
    \includegraphics[width=1.0\linewidth, angle=0]{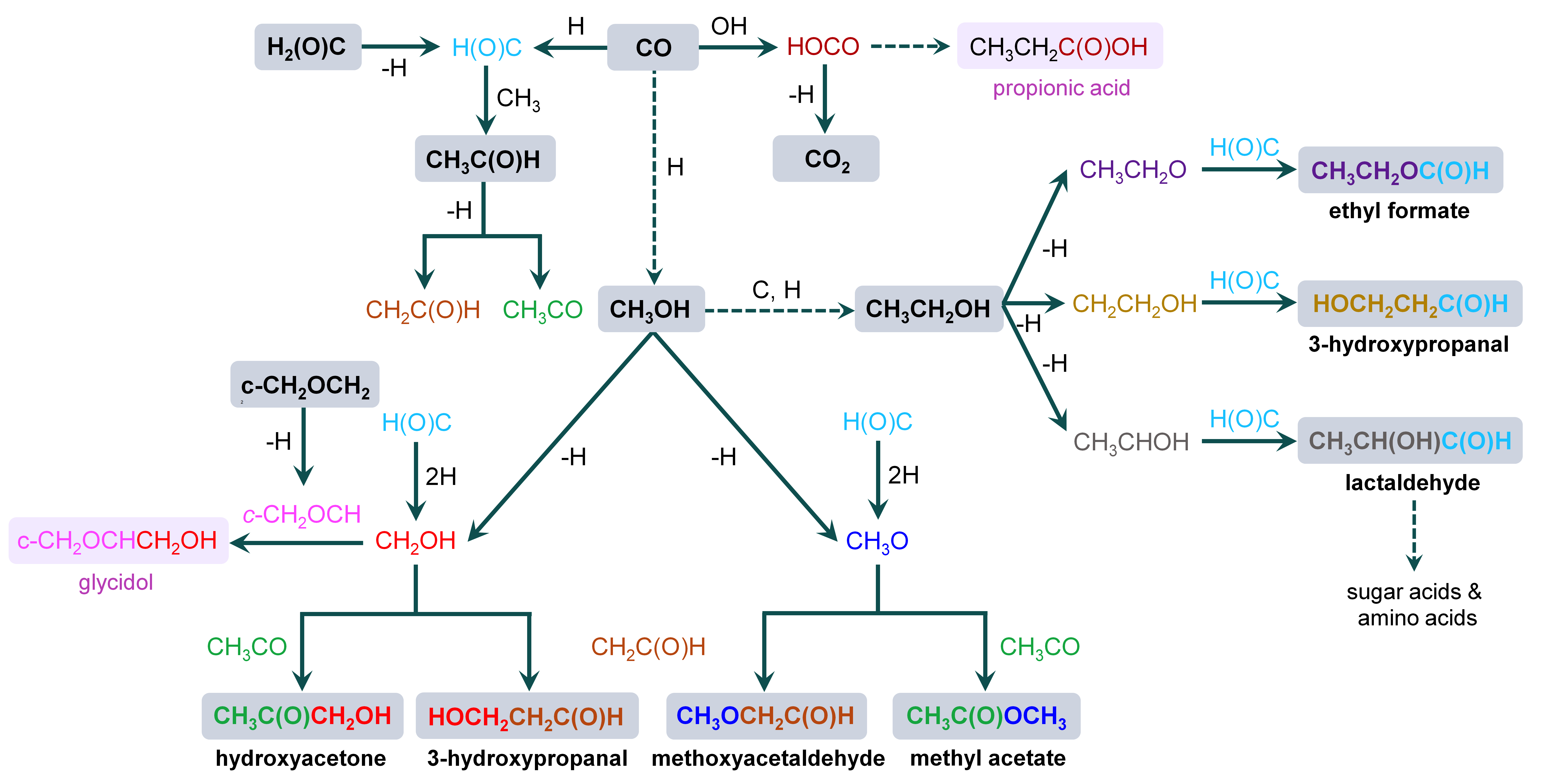}
    \caption{Suggested grain-surface chemical routes for the formation of the targeted C$_3$H$_6$O$_2$ isomers in the ISM. Dashed lines indicated processes involving more than one step. We highlight in boldface molecules that have been identified toward G+0.693. C$_3$H$_6$O$_2$ species that have not been detected and whose formation has not been tested experimentally by \cite{wang23} and \cite{wang24} are highlighted with a light purple box. For simplicity, stereoisomerism is not considered in this plot.
}
\label{f:model}
\end{figure*}
\end{center}


\section{Astrochemical implications and conclusions} 
\label{sec:disc}

The detection of six distinct C$_3$H$_6$O$_2$ structural isomers toward G+0.693 (i.e., methyl acetate, hydroxyacetone, ethyl formate, 3-hydroxypropanal, lactaldehyde and methoxyacetaldehyde), together with the non-detection of propionic acid and glycidol, allows us to delve further into the formation of these species in the ISM.

Here, we focus on possible grain-surface chemical formation pathways, which have been previously demonstrated to produce multiple \ch{C3H6O2} isomers upon UV irradiation of methanol–acetaldehyde and CO–ethanol ices \citep{wang23,wang24}. As shown in Figure \ref{f:model}, all the proposed routes proceed through a final radical–radical recombination reaction starting from the parent molecule CO, which can either be hydrogenated to form methanol (CH$_3$OH; \citealt{Jimenez-Serra25}) or oxygenated to yield HOCO and, eventually, CO$_2$ \citep{Garrod2011,Minissale2013,Molpeceres2025}. In the former case, CH$_3$OH can undergo H-abstraction to produce the CH$_2$OH and CH$_3$O radicals, which may subsequently react with CH$_3$CO and CH$_2$C(O)H to form hydroxyacetone, 3-hydroxypropanal, methyl acetate and methoxyacetaldehyde, although to our knowledge the latter has not been detected in previous experiments \citep{wang23,wang24}. Alternatively, CH$_3$OH can evolve into CH$_3$CH$_2$OH, which, upon H-abstraction, yields CH$_3$CH$_2$O, CH$_2$CH$_2$OH, or CH$_3$CHOH, depending on the abstraction site. These radicals can then react with the H(O)C radical, thought to form efficiently via hydrogenation of CO \citep{Brown1988grains} or H-abstraction from formaldehyde (H$_2$(O)C), leading to the formation of ethyl formate, 3-hydroxypropanal, and lactaldehyde, respectively. Remarkably, lactaldehyde has been proposed as a key building block for relevant prebiotic molecules, such as pyruvic acid, methylglyoxal, and lactic acid, which participate in the methylglyoxal cycle and are considered as precursors of amino acids, sugars, and sugar acids \citep{wang24}. 

While the formation routes described above could indeed account for the formation of the targeted C$_3$H$_6$O$_2$ isomers in the ISM, their efficiency depends critically on two ingredients: i) a high cosmic-ray ionization rate (CRIR) to promote radical–radical recombination reactions, and ii) access to the chemistry that occurs on the surface of dust grains, enabled by the desorption of ice-mantle species into the gas phase. This aligns perfectly with two of the key factors driving the exceptionally rich chemistry observed in G+0.693. First, large-scale, low-velocity shocks, likely resulting from a cloud–cloud collision scenario \citep{requena-torres_organic_2006, zeng2020, Colzi2024}, enhance the sputtering of icy grain mantles, thereby exposing surface chemistry. Second, the high CRIR that is thought to affect the cloud (10$^{-14}$–10$^{-15}$ s$^{-1}$), inferred from chemical modeling of the PO$^+$ and HOCS$^+$ cations \citep{Rivilla2022c, Sanz-Novo2024a}, has been suggested to facilitate the formation of radicals on these icy mantles. Once generated, these radicals can efficiently interact and recombine, ultimately driving the synthesis of increasingly complex molecules \citep{Rivilla23}.

On a different note, the non-detection of propionic acid and glycidol provides additional valuable information. These species exhibit the lowest dipole moments among all isomers (see Figure \ref{f:energies}), which significantly hampers their astronomical detection. Interestingly, while most C$_3$H$_6$O$_2$ isomers can be formed through the hydrogenation of CO (Figure \ref{f:model}), propionic acid follows a distinct chemical route. Its formation begins with the oxygenation of CO via the reaction CO + OH, producing the HOCO radical, which is widely considered a key precursor in the formation of interstellar carboxylic acids (see, e.g., \citealt{Oba2010,Ioppolo2021,Ishibashi2024,Molpeceres2025}). This species can either be converted into CO or CO$_2$ \citep{Molpeceres2025} or, alternatively, it could lead to the formation of more complex acids such as propionic acid. However, this latter pathway might represent only a minor channel, given the limited reservoir of complex carboxylic acids currently known in the ISM \citep{Sanz-Novo2023}. Furthermore, the chemical distinction proposed here between propionic acid and the other C$_3$H$_6$O$_2$ isomers is consistent with the spatial differentiation observed in ALMA interferometric observations toward Orion KL by \cite{Tercero2018}, which is clearly differentiated for methyl acetate and ethyl formate compared to that of the carboxylic acids (HCOOH and CH$_3$COOH). 

At this point, we can compare the upper limit to the column density derived for propionic acid with the abundance of the related $trans$-formic acid ($t$-HCOOH; $N$ = (2.0 $\pm$ 0.4)  $\times$ 10$^{14}$ cm$^{-2}$) and acetic acid (CH$_3$COOH; $N$ = (4.5 $\pm$ 0.2) $\times$ 10$^{13}$ cm$^{-2}$), both of which are also detected toward G+0.693 \citep{Sanz-Novo2023}. As has been observed for other chemical families such as alcohols, thiols, aldehydes, and isocyanates \citep{rodriguez-almeida2021a,rodriguez-almeida2021b,Jimenez-Serra22,sanz-novo2022,Fried2025}, the abundances typically decrease by roughly one order of magnitude with increasing molecular complexity (i.e., the substitution of a –CH$_3$ group). In this case, the observed ratios are $t$-HCOOH/CH$_3$COOH = 4.4 $\pm$ 0.9 and CH$_3$COOH/CH$_3$CH$_2$COOH $\geq$2.0, which suggests that we may still be below the required sensitivity to detect CH$_3$CH$_2$COOH. 

Regarding glycidol, while the proposed formation route also begins with CH$_3$OH, it requires an additional precursor for the suggested radical–radical recombination reaction: $c$-CH$_2$OCH. This radical could be produced via H-abstraction of ethylene oxide ($c$-CH$_2$OCH$_2$). However, the relatively low abundance of $c$-CH$_2$OCH$_2$ observed toward G+0.693 ($N$ $\sim$ 10$^{14}$ cm$^{-2}$; fractional abundance of $\sim$7 $\times$ 10$^{-10}$; \citealt{requena-torres_largest_2008}) compared with the other precursors such as CH$_3$OH and CH$_3$CH$_2$OH, which are $\sim$2 and $\sim$1 orders of magnitude more abundant than $c$-CH$_2$OCH$_2$, respectively (i.e., 1.1$\pm$0.2 $\times$ 10$^{-7}$ and 4.6$\pm$0.6 $\times$ 10$^{-9}$; \citealt{rodriguez-almeida2021a}), may hamper the detection of glycidol in the present survey. Additionally, we highlight that neither glycidol nor propionic acid were found in the experiments by \cite{wang23,wang24}.

\begin{figure*}
\centerline{\resizebox{1\hsize}{!}{\includegraphics[angle=0]{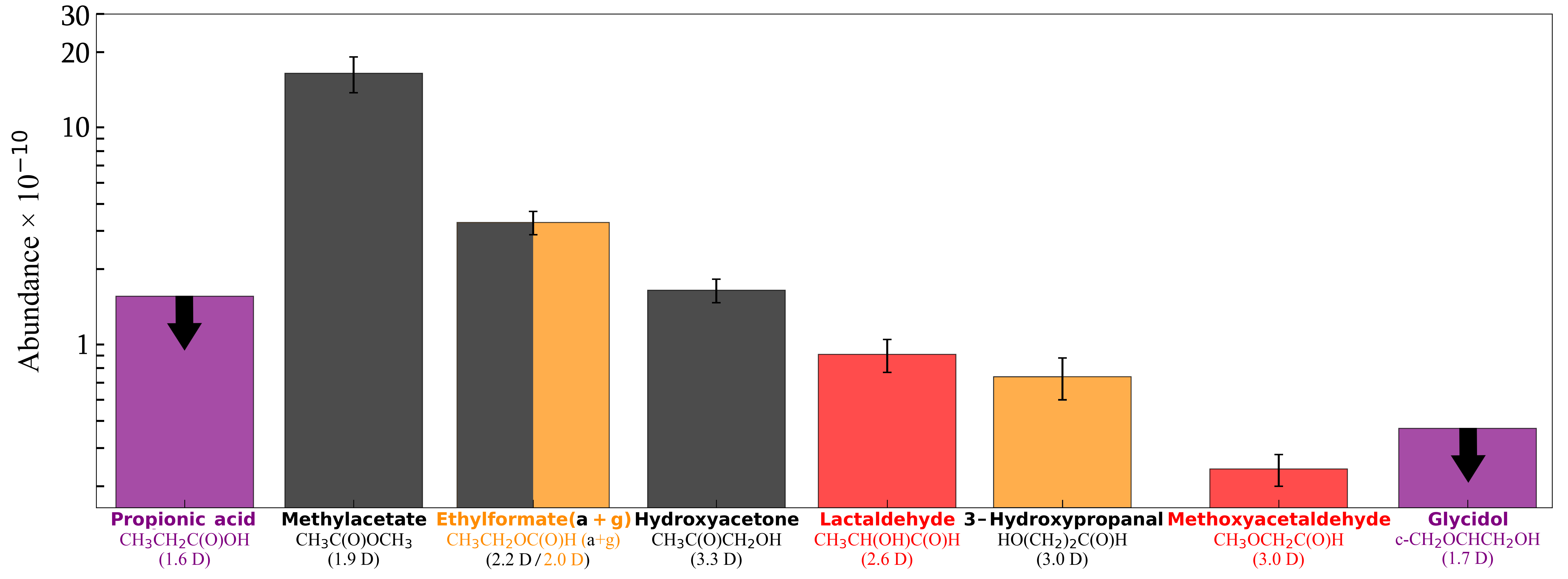}}}
\caption{Fractional abundances with respect to H$_2$ of the different C$_3$H$_6$O$_2$ isomers sorted by increasing relative energy. The total dipole moment for each molecule (computed at the B2PLYPD3/aug-cc-pVTZ level) is included. We use the same color code as that in Figure \ref{f:energies}: new first detections in the ISM are shown in red, tentative detections in orange, non-detections in purple (upper limits indicated with black arrows) and previous detections toward different sources in black.}
\label{f:abundances}
\end{figure*}

Finally, the comparison of the abundances of the detected C$_3$H$_6$O$_2$ isomers (Table \ref{tab:comparison}) leads to the following ordering (see Figure \ref{f:abundances}):

$N$(methyl acetate) $>$$~$$N$(ethyl formate) $>$ $N$(hydroxyacetone) $>$  $N$(lactaldehyde) $>$ $N$(3-hydroxypropanal) $>$$~$$N$(methoxyacetaldehyde)

Methyl acetate clearly emerges as the dominant isomer within the C$_3$H$_6$O$_2$ family, followed by ethyl formate and hydroxyacetone. Surprisingly, the observed abundance trend qualitatively follows the relative thermodynamic stability of the detected isomers (Figure \ref{f:abundances}). However, a major exception is propionic acid, the most stable member of the family, which remains undetected. This discrepancy shows that thermodynamic stability alone cannot be used as a reliable tool to predict the molecular abundances within this isomeric family in the ISM, as reported for other families (e.g., C$_3$H$_2$O, C$_2$H$_4$O$_2$, C$_2$H$_2$N$_2$, and C$_2$H$_5$NO$_2$; \citealt{Loomis:2015jh, Mininni2020, Shingledecker2020, Rivilla23, SanAndres2024}). Instead, kinetics are likely responsible for discriminating between species arising from separated chemical formation pathways (i.e., those originating from CH$_3$OH or from HOCO; see Figure \ref{f:model}).

On the other hand, the significantly higher abundance of methyl acetate may suggest that the formation of this species is more efficient, potentially due to additional formation pathways occurring either on dust-grains or in the gas-phase (e.g., \citealt{Das2015}). Alternatively, if the relative dipole principle (RDP; \citealt{Shingledecker2020}) applies to this isomeric family, then the lower dipole moment of methyl acetate ($\mu$$_{tot}$ = 1.9 D, see Table \ref{tab:comparison}) would lead to a slower destruction in the gas phase compared to the other detected species. Nevertheless, further theoretical, laboratory and modeling effort is needed to shed conclusive light on this question.

In summary, our systematic survey of all C$_3$H$_6$O$_2$ isomers toward G+0.693 has significantly expanded the census of complex organic molecules in the ISM, particularly within this isomeric family. We report the interstellar discovery of lactaldehyde and methoxyacetaldehyde, along with the confirmation of methyl acetate and hydroxyacetone, and new detections in G+0.693 of both $anti$- and $gauche$-conformers of ethyl formate. These results not only enhance our understanding of the molecular inventory in the ISM but also emphasize the importance of highly sensitive spectral line surveys for detecting the less abundant species. Furthermore, they provide a rationale for the more than decade-long gap between the detection of methyl acetate and ethyl formate and the new interstellar discoveries reported in this work. These results also underscore the need for future laboratory and observational efforts targeting other uncharacterized C$_3$H$_6$O$_2$ isomers, including various enol tautomers (e.g., prop-1-ene-1,2-diol, prop-2-ene-1,2-diol, 1-methoxyethen-1-ol, prop-1-ene-1,3-diol) proposed by \cite{wang23}, which may be efficiently produced via keto–enol tautomerization. Based on our findings, these species also appear as promising candidates for future interstellar detection.


\begin{acknowledgements}
 
We are grateful to the IRAM 30$\,$m and Yebes 40$\,$m telescopes staff for their help during the different observing runs, highlighting project 21A014 (PI: Rivilla), project 018-19 (PI: Rivilla) and projects 123-22 and 076-23 (PI: Jim\'enez-Serra). The 40$\,$m radio telescope at Yebes Observatory is operated by the Spanish Geographic Institute (IGN, Ministerio de Transportes, Movilidad y Agenda Urbana). IRAM is supported by INSU/CNRS (France), MPG (Germany) and IGN (Spain). M.S.-N. acknowledges a Juan de la Cierva Postdoctoral Fellow proyect JDC2022-048934-I, funded by the Spanish Ministry of Science, Innovation and Universities/State Agency of Research MICIU/AEI/10.13039/501100011033 and by the European Union “NextGenerationEU”/PRTR”. V.M.R. acknowledges support from the grant RYC2020-029387-I funded by MICIU/AEI/10.13039/501100011033 and by “ESF, Investing in your future". V.M.R., A.L-G. and D.S.A from the Consejo Superior de Investigaciones Cient{\'i}ficas (CSIC) and the Centro de Astrobiolog{\'i}a (CAB) through the project 20225AT015 (Proyectos intramurales especiales del CSIC), and from the grant CNS2023-144464 funded by MICIU/AEI/10.13039/501100011033 and by “European Union NextGenerationEU/PRTR”. I.J.-S., V.M.R., M.S.-N, L.C., A.M., D.S.A, A.L-G. and A.M.-H. acknowledge funding from grant No. PID2022-136814NB-I00 from MICIU/AEI/10.13039/501100011033 and by “ERDF, UE”. I.J.-S. and A.M. also acknowledge funding from the ERC grant OPENS (project number 101125858) funded by the European Union. Views and opinions expressed are however those of the author(s) only and do not necessarily reflect those of the European Union or the European Research Council Executive Agency. Neither the European Union nor the granting authority can be held responsible for them. D.S.A. also extends his gratitude for the financial support provided by the Comunidad de Madrid through the Grant PIPF-2022/TEC-25475. A.M.-H. acknowledges funding from grant MDM-2017-0737 Unidad de Excelencia “María de Maeztu” Centro de Astrobiología (CAB, CSIC-INTA) funded by MICIU/AEI/10.13039/501100011033. B.A.M. and Z.T.P.F. acknowledge support from Schmidt Family Futures. L.K. acknowledges financial funding from the Czech Science Foundation (GACR, grant No. 24-12586S). B.T. acknowledges Spanish Ministry of Science support from grants PID2022-137980NB-100 and PID2023-147545NB-I00. J.-C.G. thanks the Centre National d’Etudes Spatiales (CNES) for a grant and the Programme National “Physique et Chimie du Milieu Interstellaire” (PCMI) of CNRS/INSU with INC/INP co-funded by CEA and CNES. E.J.C. acknowledges financial support from the Spanish Ministry of Science and Innovation (MCIN/AEI, Project PID2023-147698NB-I00) and the Basque Government (Project IT1491-22) I.K. would like to thank the Agence Nationale de la recherche Scientifique (ANR) project ANR-25-CE29-7779. 

\end{acknowledgements}

\bibliography{rivilla,bibliography}{}

@ARTICLE{Nguyen2014,
       author = {{Nguyen}, Ha Vinh Lam and {Kleiner}, Isabelle and {Shipman}, Steven T. and {Mae}, Yoshiaki and {Hirose}, Kazue and {Hatanaka}, Shota and {Kobayashi}, Kaori},
        title = "{Extension of the measurement, assignment, and fit of the rotational spectrum of the two-top molecule methyl acetate}",
      journal = {Journal of Molecular Spectroscopy},
         year = 2014,
        month = may,
       volume = {299},
        pages = {17-21},
          doi = {10.1016/j.jms.2014.03.012},
       adsurl = {https://ui.adsabs.harvard.edu/abs/2014JMoSp.299...17N}
}

@ARTICLE{Carvajal2019,
       author = {{Carvajal}, M. and {Favre}, C. and {Kleiner}, I. and {Ceccarelli}, C. and {Bergin}, E.~A. and {Fedele}, D.},
        title = "{Impact of nonconvergence and various approximations of the partition function on the molecular column densities in the interstellar medium}",
      journal = {\aap},
         year = 2019,
        month = jul,
       volume = {627},
          eid = {A65},
        pages = {A65},
          doi = {10.1051/0004-6361/201935469},
archivePrefix = {arXiv},
       eprint = {1906.02067},
 primaryClass = {astro-ph.GA},
       adsurl = {https://ui.adsabs.harvard.edu/abs/2019A&A...627A..65C}
}

@ARTICLE{Sanz-Novo2025b,
       author = {{Sanz-Novo}, M. and {Molpeceres}, G. and {Rivilla}, V.~M. and {Jimenez-Serra}, I.},
        title = "{Conformational isomerism of methyl formate: New detections of the higher-energy trans conformer and theoretical insights}",
      journal = {\aap},
         year = 2025,
        month = may,
       volume = {698},
          eid = {A36},
        pages = {A36},
          doi = {10.1051/0004-6361/202554686},
archivePrefix = {arXiv},
       eprint = {2503.23045},
 primaryClass = {astro-ph.GA},
       adsurl = {https://ui.adsabs.harvard.edu/abs/2025A&A...698A..36S}
}

@article{peng_alma_2019,
	title = {{ALMA} {Observations} of {Ethyl} {Formate} toward {Orion} {KL}},
	volume = {871},
	issn = {0004-637X},
	url = {https://dx.doi.org/10.3847/1538-4357/aafad4},
	doi = {10.3847/1538-4357/aafad4},
	language = {en},
	number = {2},
	urldate = {2025-02-28},
	journal = {\apj},
	author = {Peng, Yaping and Rivilla, V. M. and Zhang, Li and Ge, J. X. and Zhou, Bing},
	month = feb,
	year = {2019},
	pages = {251},
}

@article{ethyl_formate_Belloche,
	title = {Increased complexity in interstellar chemistry: detection and chemical modeling of ethyl formate and n-propyl cyanide in {Sagittarius} {B2}({N})},
	volume = {499},
	copyright = {© ESO, 2009},
	issn = {0004-6361, 1432-0746},
	shorttitle = {Increased complexity in interstellar chemistry},
	url = {https://www.aanda.org/articles/aa/abs/2009/19/aa11550-08/aa11550-08.html},
	doi = {10.1051/0004-6361/200811550},
	language = {en},
	number = {1},
	urldate = {2024-04-25},
	journal = {Astronomy \& Astrophysics},
	author = {Belloche, A. and Garrod, R. T. and Müller, H. S. P. and Menten, K. M. and Comito, C. and Schilke, P.},
	month = may,
	year = {2009},
	note = {Number: 1
Publisher: EDP Sciences},
	pages = {215--232},
}

@article{methyl_acetate_Orion,
	title = {{DISCOVERY} {OF} {METHYL} {ACETATE} {AND} {GAUCHE} {ETHYL} {FORMATE} {IN} {ORION}*},
	volume = {770},
	issn = {2041-8205},
	url = {https://dx.doi.org/10.1088/2041-8205/770/1/L13},
	doi = {10.1088/2041-8205/770/1/L13},
	language = {en},
	number = {1},
	urldate = {2024-04-25},
	journal = {The Astrophysical Journal Letters},
	author = {Tercero, B. and Kleiner, I. and Cernicharo, J. and Nguyen, H. V. L. and López, A. and Caro, G. M. Muñoz},
	month = may,
	year = {2013},
	pages = {L13},
}

@article{hydroxyacetone_IRAS,
	title = {Detection of hydroxyacetone in protostar {IRAS} 16293-2422 {B}},
	volume = {20},
	issn = {1674-4527},
	url = {https://dx.doi.org/10.1088/1674-4527/20/8/125},
	doi = {10.1088/1674-4527/20/8/125},
	language = {en},
	number = {8},
	urldate = {2024-04-25},
	journal = {Research in Astronomy and Astrophysics},
	author = {Zhou, Yan and Quan, Dong-Hui and Zhang, Xia and Qin, Sheng-Li},
	month = aug,
	year = {2020},
	pages = {125},
}

@article{methoxyacetaldehyde_Kolesnikova,
	title = {Laboratory rotational spectrum and astronomical search for methoxyacetaldehyde},
	volume = {619},
	copyright = {© ESO 2018},
	issn = {0004-6361, 1432-0746},
	url = {https://www.aanda.org/articles/aa/abs/2018/11/aa33773-18/aa33773-18.html},
	doi = {10.1051/0004-6361/201833773},
	language = {en},
	urldate = {2024-04-25},
	journal = {Astronomy \& Astrophysics},
	author = {Kolesniková, L. and Peña, I. and Alonso, E. R. and Tercero, B. and Cernicharo, J. and Mata, S. and Alonso, J. L.},
	month = nov,
	year = {2018},
	pages = {A67},
}

@article{lactaldehyde_Alonso,
	title = {The {Laboratory} {Millimeter} and {Submillimeter} {Rotational} {Spectrum} of {Lactaldehyde} and an {Astronomical} {Search} in {Sgr} {B2}({N}), {Orion}-{KL}, and {NGC} {6334I}},
	volume = {883},
	issn = {0004-637X},
	url = {https://dx.doi.org/10.3847/1538-4357/ab3463},
	doi = {10.3847/1538-4357/ab3463},
	language = {en},
	number = {1},
	urldate = {2024-04-25},
	journal = {\apj},
	author = {Alonso, Elena R. and McGuire, Brett A. and Kolesniková, Lucie and Carroll, P. Brandon and León, Iker and Brogan, Crystal L. and Hunter, Todd R. and Guillemin, Jean-Claude and Alonso, Jose L.},
	month = sep,
	year = {2019},
	pages = {18},
}

@article{ilyushin_submillimeter_2021,
	title = {Submillimeter wave spectroscopy of propanoic acid ({CH3CH2COOH}) and its {ISM} search},
	volume = {379},
	issn = {0022-2852},
	url = {https://www.sciencedirect.com/science/article/pii/S0022285221000382},
	doi = {10.1016/j.jms.2021.111454},
	urldate = {2025-02-28},
	journal = {Journal of Molecular Spectroscopy},
	author = {Ilyushin, V. V. and Margulès, L. and Tercero, B. and Motiyenko, R. A. and Dorovskaya, O. and Alekseev, E. A. and Alonso, E. R. and Kolesniková, L. and Cernicharo, J. and Guillemin, J. C.},
	month = may,
	year = {2021},
	pages = {111454},
}

@ARTICLE{Remijan2025,
       author = {{Remijan}, Anthony J. and {Changala}, P. Bryan and {Xue}, Ci and {Yuan}, Elsa Q.~H. and {Duffy}, Miya and {Scolati}, Haley N. and {Shingledecker}, Christopher N. and {Speak}, Thomas H. and {Cooke}, Ilsa R. and {Loomis}, Ryan and {Burkhardt}, Andrew M. and {Fried}, Zachary T.~P. and {Wenzel}, Gabi and {Lipnicky}, Andrew and {McCarthy}, Michael C. and {McGuire}, Brett A.},
        title = "{The Missing Link of Sulfur Chemistry in TMC-1: The Detection of c-C$_{3}$H$_{2}$S from the GOTHAM Survey}",
      journal = {\apj},
         year = 2025,
        month = apr,
       volume = {982},
       number = {2},
          eid = {191},
        pages = {191},
          doi = {10.3847/1538-4357/adb84c},
archivePrefix = {arXiv},
       eprint = {2501.06343},
 primaryClass = {astro-ph.GA},
       adsurl = {https://ui.adsabs.harvard.edu/abs/2025ApJ...982..191R}
}

@ARTICLE{Marcelino2021,
       author = {{Marcelino}, N. and {Tercero}, B. and {Ag{\'u}ndez}, M. and {Cernicharo}, J.},
        title = "{A study of C$_{4}$H$_{3}$N isomers in TMC-1: Line by line detection of HCCCH$_{2}$CN}",
      journal = {\aap},
         year = 2021,
        month = feb,
       volume = {646},
          eid = {L9},
        pages = {L9},
          doi = {10.1051/0004-6361/202040177},
archivePrefix = {arXiv},
       eprint = {2101.08516},
 primaryClass = {astro-ph.GA},
       adsurl = {https://ui.adsabs.harvard.edu/abs/2021A&A...646L...9M}
}

@ARTICLE{Garcia_de_la_concepcion2023,
       author = {{Garc{\'\i}a de la Concepci{\'o}n}, J. and {Jim{\'e}nez-Serra}, I. and {Corchado}, J.~C. and {Molpeceres}, G. and {Mart{\'\i}nez-Henares}, A. and {Rivilla}, V.~M. and {Colzi}, L. and {Mart{\'\i}n-Pintado}, J.},
        title = "{A sequential acid-base mechanism in the interstellar medium: The emergence of cis-formic acid in dark molecular clouds}",
      journal = {\aap},
         year = 2023,
        month = jul,
       volume = {675},
          eid = {A109},
        pages = {A109},
          doi = {10.1051/0004-6361/202243966},
archivePrefix = {arXiv},
       eprint = {2301.07450},
 primaryClass = {astro-ph.GA},
       adsurl = {https://ui.adsabs.harvard.edu/abs/2023A&A...675A.109G}
}

@ARTICLE{Sanz-Novo2023,
       author = {{Sanz-Novo}, Miguel and {Rivilla}, V{\'\i}ctor M. and {Jim{\'e}nez-Serra}, Izaskun and {Mart{\'\i}n-Pintado}, Jes{\'u}s and {Colzi}, Laura and {Zeng}, Shaoshan and {Meg{\'\i}as}, Andr{\'e}s and {L{\'o}pez-Gallifa}, {\'A}lvaro and {Mart{\'\i}nez-Henares}, Antonio and {Massalkhi}, Sarah and {Tercero}, Bel{\'e}n and {de Vicente}, Pablo and {Mart{\'\i}n}, Sergio and {San Andr{\'e}s}, David and {Requena-Torres}, Miguel A.},
        title = "{Discovery of the Elusive Carbonic Acid (HOCOOH) in Space}",
      journal = {\apj},
         year = 2023,
        month = sep,
       volume = {954},
       number = {1},
          eid = {3},
        pages = {3},
          doi = {10.3847/1538-4357/ace523},
archivePrefix = {arXiv},
       eprint = {2307.08644},
 primaryClass = {astro-ph.GA},
       adsurl = {https://ui.adsabs.harvard.edu/abs/2023ApJ...954....3S}
}

@article{Sanz-Novo2025a,
doi = {10.3847/2041-8213/adafa7},
url = {https://dx.doi.org/10.3847/2041-8213/adafa7},
year = {2025},
month = {feb},
publisher = {The American Astronomical Society},
volume = {980},
number = {2},
pages = {L37},
author = {Sanz-Novo, Miguel and Rivilla, Víctor M. and Endres, Christian P. and Lattanzi, Valerio and Jiménez-Serra, Izaskun and Colzi, Laura and Zeng, Shaoshan and Megías, Andrés and López-Gallifa, Álvaro and Martínez-Henares, Antonio and San Andrés, David and Tercero, Belén and de Vicente, Pablo and Martín, Sergio and Requena-Torres, Miguel A. and Caselli, Paola and Martín-Pintado, Jesús},
title = {On the Abiotic Origin of Dimethyl Sulfide: Discovery of Dimethyl Sulfide in the Interstellar Medium},
journal = {The Astrophysical Journal Letters}
}

@ARTICLE{wang24,
       author = {{Wang}, Jia and {Zhang}, Chaojiang and {Marks}, Joshua H. and {Evseev}, Mikhail M. and {Kuznetsov}, Oleg V. and {Antonov}, Ivan O. and {Kaiser}, Ralf I.},
        title = "{Interstellar formation of lactaldehyde, a key intermediate in the methylglyoxal pathway}",
      journal = {Nature Communications},
         year = 2024,
        month = nov,
       volume = {15},
       number = {1},
          eid = {10189},
        pages = {10189},
          doi = {10.1038/s41467-024-54562-x},
       adsurl = {https://ui.adsabs.harvard.edu/abs/2024NatCo..1510189W}
}

@article{wang23,
	title = {Mechanistical study on the formation of hydroxyacetone ({CH3COCH2OH}), methyl acetate ({CH3COOCH3}), and 3-hydroxypropanal ({HCOCH2CH2OH}) along with their enol tautomers (prop-1-ene-1,2-diol ({CH3C}({OH}){CHOH}), prop-2-ene-1,2-diol ({CH2C}({OH}){CH2OH}), 1-methoxyethen-1-ol ({CH3OC}({OH}){CH2}) and prop-1-ene-1,3-diol ({HOCH2CHCHOH})) in interstellar ice analogs},
	volume = {25},
	issn = {1463-9084},
	url = {https://pubs.rsc.org/en/content/articlelanding/2023/cp/d2cp03543j},
	doi = {10.1039/D2CP03543J},
	language = {en},
	number = {2},
	urldate = {2024-04-25},
	journal = {Physical Chemistry Chemical Physics},
	author = {Wang, Jia and Marks, Joshua H. and Turner, Andrew M. and Nikolayev, Anatoliy A. and Azyazov, Valeriy and Mebel, Alexander M. and Kaiser, Ralf I.},
	month = jan,
	year = {2023},
	pages = {936--953},
}

@ARTICLE{Fried2025,
doi = {10.3847/1538-4357/adfc62},
url = {https://doi.org/10.3847/1538-4357/adfc62},
year = {2025},
month = {oct},
publisher = {The American Astronomical Society},
volume = {992},
number = {2},
pages = {187},
author = {Fried, Zachary T. P. and Motiyenko, Roman A. and Sanz-Novo, Miguel and Kolesniková, Lucie and Guillemin, Jean-Claude and Margulès, Laurent and Uhlíková, Tereza and Belloche, Arnaud and Jørgensen, Jes K. and Holdren, Martin S. and Xue, Ci and Urban, Štěpán and Jiménez-Serra, Izaskun and Rivilla, Victor M. and McGuire, Brett A.},
title = {Rotational Spectroscopy and Tentative Interstellar Detection of 3-hydroxypropanal (HOCH2CH2CHO) in the G+0.693-0.027 Molecular Cloud},
journal = {\apj}
}

@ARTICLE{Belloche2025,
       author = {{Belloche}, A. and {Garrod}, R.~T. and {M{\"u}ller}, H.~S.~P. and {Morin}, N.~J. and {Willis}, S.~A. and {Menten}, K.~M.},
        title = "{Re-exploring Molecular Complexity with ALMA: Insights into chemical differentiation from the molecular composition of hot cores in Sgr B2(N2)}",
      journal = {\aap},
         year = 2025,
        month = jun,
       volume = {698},
          eid = {A143},
        pages = {A143},
          doi = {10.1051/0004-6361/202554411},
archivePrefix = {arXiv},
       eprint = {2505.03262},
 primaryClass = {astro-ph.GA},
       adsurl = {https://ui.adsabs.harvard.edu/abs/2025A&A...698A.143B}
}

@misc{Frisch2016,
author={M. J. Frisch and G. W. Trucks and H. B. Schlegel and G. E. Scuseria and M. A. Robb and J. R. Cheeseman and G. Scalmani and V. Barone and G. A. Petersson and H. Nakatsuji and X. Li and M. Caricato and A. V. Marenich and J. Bloino and B. G. Janesko and R. Gomperts and B. Mennucci and H. P. Hratchian and J. V. Ortiz and A. F. Izmaylov and J. L. Sonnenberg and D. Williams-Young and F. Ding and F. Lipparini and F. Egidi and J. Goings and B. Peng and A. Petrone and T. Henderson and D. Ranasinghe and V. G. Zakrzewski and J. Gao and N. Rega and G. Zheng and W. Liang and M. Hada and M. Ehara and K. Toyota and R. Fukuda and J. Hasegawa and M. Ishida and T. Nakajima and Y. Honda and O. Kitao and H. Nakai and T. Vreven and K. Throssell and Montgomery, {Jr.}, J. A. and J. E. Peralta and F. Ogliaro and M. J. Bearpark and J. J. Heyd and E. N. Brothers and K. N. Kudin and V. N. Staroverov and T. A. Keith and R. Kobayashi and J. Normand and K. Raghavachari and A. P. Rendell and J. C. Burant and S. S. Iyengar and J. Tomasi and M. Cossi and J. M. Millam and M. Klene and C. Adamo and R. Cammi and J. W. Ochterski and R. L. Martin and K. Morokuma and O. Farkas and J. B. Foresman and D. J. Fox},
title={Gaussian˜16 {R}evision {C}.01},
year={2016},
note={Gaussian Inc. Wallingford CT}
}

@ARTICLE{Colzi2024,
       author = {{Colzi}, L. and {Mart{\'\i}n-Pintado}, J. and {Zeng}, S. and {Jim{\'e}nez-Serra}, I. and {Rivilla}, V.~M. and {Sanz-Novo}, M. and {Mart{\'\i}n}, S. and {Zhang}, Q. and {Lu}, X.},
        title = "{Excitation and spatial study of a prestellar cluster towards G+0.693-0.027 in the Galactic centre}",
      journal = {\aap},
         year = 2024,
        month = oct,
       volume = {690},
          eid = {A121},
        pages = {A121},
          doi = {10.1051/0004-6361/202451382},
archivePrefix = {arXiv},
       eprint = {2408.17141},
 primaryClass = {astro-ph.GA},
       adsurl = {https://ui.adsabs.harvard.edu/abs/2024A&A...690A.121C}
}

@ARTICLE{SanAndres2024,
       author = {{San Andr{\'e}s}, David and {Rivilla}, V{\'\i}ctor M. and {Colzi}, Laura and {Jim{\'e}nez-Serra}, Izaskun and {Mart{\'\i}n-Pintado}, Jes{\'u}s and {Meg{\'\i}as}, Andr{\'e}s and {L{\'o}pez-Gallifa}, {\'A}lvaro and {Mart{\'\i}nez-Henares}, Antonio and {Massalkhi}, Sarah and {Zeng}, Shaoshan and {Sanz-Novo}, Miguel and {Tercero}, Bel{\'e}n and {de Vicente}, Pablo and {Mart{\'\i}n}, Sergio and {Requena Torres}, Miguel Angel and {Molpeceres}, Germ{\'a}n and {Garc{\'\i}a de la Concepci{\'o}n}, Juan},
        title = "{First Detection in Space of the High-energy Isomer of Cyanomethanimine: H$_{2}$CNCN}",
      journal = {\apj},
         year = 2024,
        month = may,
       volume = {967},
       number = {1},
          eid = {39},
        pages = {39},
          doi = {10.3847/1538-4357/ad3af3},
archivePrefix = {arXiv},
       eprint = {2404.03334},
 primaryClass = {astro-ph.GA},
       adsurl = {https://ui.adsabs.harvard.edu/abs/2024ApJ...967...39S}
}

@ARTICLE{Sanz-Novo2024a,
       author = {{Sanz-Novo}, Miguel and {Rivilla}, V{\'\i}ctor M. and {Jim{\'e}nez-Serra}, Izaskun and {Mart{\'\i}n-Pintado}, Jes{\'u}s and {Colzi}, Laura and {Zeng}, Shaoshan and {Meg{\'\i}as}, Andr{\'e}s and {L{\'o}pez-Gallifa}, {\'A}lvaro and {Mart{\'\i}nez-Henares}, Antonio and {Massalkhi}, Sarah and {Tercero}, Bel{\'e}n and {de Vicente}, Pablo and {San Andr{\'e}s}, David and {Mart{\'\i}n}, Sergio and {Requena-Torres}, Miguel A.},
        title = "{Interstellar Detection of O-protonated Carbonyl Sulfide, HOCS$^{+}$}",
      journal = {\apj},
         year = 2024,
        month = apr,
       volume = {965},
       number = {2},
          eid = {149},
        pages = {149},
          doi = {10.3847/1538-4357/ad2c01},
archivePrefix = {arXiv},
       eprint = {2402.15405},
 primaryClass = {astro-ph.GA},
       adsurl = {https://ui.adsabs.harvard.edu/abs/2024ApJ...965..149S}
}

@article{Zheng2024,
doi = {10.3847/1538-4357/ad072c},
url = {https://dx.doi.org/10.3847/1538-4357/ad072c},
year = {2024},
month = {jan},
publisher = {The American Astronomical Society},
volume = {961},
number = {1},
pages = {58},
author = {Siqi Zheng and Juan Li and Junzhi Wang and Yao Wang and Feng Gao and Donghui Quan and Fujun Du and Yajun Wu and Edwin Bergin and Yuqiang Li},
title = {Mapping Observations of Peptide-like Molecules around Sagittarius B2},
journal = {\apj}
}

@article{Li2020,
    author = {Li, Juan and Wang, Junzhi and Qiao, Haihua and Quan, Donghui and Fang, Min and Du, Fujun and Li, Fei and Shen, Zhiqiang and Li, Shanghuo and Li, Di and Shi, Yong and Zhang, Zhiyu and Zhang, Jiangshui},
    title = "{Mapping observations of complex organic molecules around Sagittarius B2 with the ARO 12 m telescope}",
    journal = {\mnras},
    volume = {492},
    number = {1},
    pages = {556-565},
    year = {2020},
    month = {01},
    issn = {0035-8711},
    doi = {10.1093/mnras/stz3337},
    url = {https://doi.org/10.1093/mnras/stz3337},
    eprint = {https://academic.oup.com/mnras/article-pdf/492/1/556/31709853/stz3337.pdf},
}

@ARTICLE{Jones2012,
       author = {{Jones}, P.~A. and {Burton}, M.~G. and {Cunningham}, M.~R. and {Requena-Torres}, M.~A. and {Menten}, K.~M. and {Schilke}, P. and {Belloche}, A. and {Leurini}, S. and {Mart{\'\i}n-Pintado}, J. and {Ott}, J. and {Walsh}, A.~J.},
        title = "{Spectral imaging of the Central Molecular Zone in multiple 3-mm molecular lines}",
      journal = {\mnras},
         year = 2012,
        month = feb,
       volume = {419},
       number = {4},
        pages = {2961-2986},
          doi = {10.1111/j.1365-2966.2011.19941.x},
archivePrefix = {arXiv},
       eprint = {1110.1421},
 primaryClass = {astro-ph.GA},
       adsurl = {https://ui.adsabs.harvard.edu/abs/2012MNRAS.419.2961J}
}

@article{Shingledecker2020,
doi = {10.3847/1538-4357/ab5360},
url = {https://dx.doi.org/10.3847/1538-4357/ab5360},
year = {2020},
month = {jan},
publisher = {The American Astronomical Society},
volume = {888},
number = {1},
pages = {52},
author = {Christopher N. Shingledecker and Thanja Lamberts and Jacob C. Laas and Anton Vasyunin and Eric Herbst and Johannes Kästner and Paola Caselli},
title = {Efficient Production of S8 in Interstellar Ices: The Effects of Cosmic-Ray-driven Radiation Chemistry and Nondiffusive Bulk Reactions},
journal = {\apj}
}

@ARTICLE{Rivilla23,
       author = {{Rivilla}, V{\'\i}ctor M. and {Sanz-Novo}, Miguel and {Jim{\'e}nez-Serra}, Izaskun and {Mart{\'\i}n-Pintado}, Jes{\'u}s and {Colzi}, Laura and {Zeng}, Shaoshan and {Meg{\'\i}as}, Andr{\'e}s and {L{\'o}pez-Gallifa}, {\'A}lvaro and {Mart{\'\i}nez-Henares}, Antonio and {Massalkhi}, Sarah and {Tercero}, Bel{\'e}n and {de Vicente}, Pablo and {Mart{\'\i}n}, Sergio and {San Andr{\'e}s}, David and {Requena-Torres}, Miguel A. and {Alonso}, Jos{\'e} Luis},
        title = "{First Glycine Isomer Detected in the Interstellar Medium: Glycolamide (NH$_{2}$C(O)CH$_{2}$OH)}",
      journal = {\apjl},
         year = 2023,
        month = aug,
       volume = {953},
       number = {2},
          eid = {L20},
        pages = {L20},
          doi = {10.3847/2041-8213/ace977},
archivePrefix = {arXiv},
       eprint = {2307.11507},
 primaryClass = {astro-ph.GA},
       adsurl = {https://ui.adsabs.harvard.edu/abs/2023ApJ...953L..20R}
}

@ARTICLE{Das2015,
       author = {{Das}, Ankan and {Majumdar}, Liton and {Sahu}, Dipen and {Gorai}, Prasanta and {Sivaraman}, B. and {Chakrabarti}, Sandip K.},
        title = "{Methyl Acetate and Its Singly Deuterated Isotopomers in the Interstellar Medium}",
      journal = {\apj},
         year = 2015,
        month = jul,
       volume = {808},
       number = {1},
          eid = {21},
        pages = {21},
          doi = {10.1088/0004-637X/808/1/21},
archivePrefix = {arXiv},
       eprint = {1504.06429},
 primaryClass = {astro-ph.GA},
       adsurl = {https://ui.adsabs.harvard.edu/abs/2015ApJ...808...21D}
}

@ARTICLE{Tercero2018,
       author = {{Tercero}, B. and {Cuadrado}, S. and {L{\'o}pez}, A. and {Brouillet}, N. and {Despois}, D. and {Cernicharo}, J.},
        title = "{Chemical segregation of complex organic O-bearing species in Orion KL}",
      journal = {\aap},
         year = 2018,
        month = nov,
       volume = {620},
          eid = {L6},
        pages = {L6},
          doi = {10.1051/0004-6361/201834417},
archivePrefix = {arXiv},
       eprint = {1811.08765},
 primaryClass = {astro-ph.GA},
       adsurl = {https://ui.adsabs.harvard.edu/abs/2018A&A...620L...6T}
}

@ARTICLE{Oba2010,
       author = {{Oba}, Yasuhiro and {Watanabe}, Naoki and {Kouchi}, Akira and {Hama}, Tetsuya and {Pirronello}, Valerio},
        title = "{Formation of Carbonic Acid (H$_{2}$CO$_{3}$) by Surface Reactions of Non-energetic OH Radicals with CO Molecules at Low Temperatures}",
      journal = {\apj},
         year = 2010,
        month = oct,
       volume = {722},
       number = {2},
        pages = {1598-1606},
          doi = {10.1088/0004-637X/722/2/1598},
       adsurl = {https://ui.adsabs.harvard.edu/abs/2010ApJ...722.1598O}
}

@ARTICLE{Ioppolo2021,
       author = {{Ioppolo}, S. and {Ka{\v{n}}uchov{\'a}}, Z. and {James}, R.~L. and {Dawes}, A. and {Ryabov}, A. and {Dezalay}, J. and {Jones}, N.~C. and {Hoffmann}, S.~V. and {Mason}, N.~J. and {Strazzulla}, G.},
        title = "{Vacuum ultraviolet photoabsorption spectroscopy of space-related ices: formation and destruction of solid carbonic acid upon 1 keV electron irradiation}",
      journal = {\aap},
         year = 2021,
        month = feb,
       volume = {646},
          eid = {A172},
        pages = {A172},
          doi = {10.1051/0004-6361/202039184},
archivePrefix = {arXiv},
       eprint = {2012.14863},
 primaryClass = {astro-ph.IM},
       adsurl = {https://ui.adsabs.harvard.edu/abs/2021A&A...646A.172I}
}

@ARTICLE{Herbst2020,
        author = {{Herbst}, Eric and {Vidali} Gianfranco and {Ceccarelli}, Cecilia},
        title = "{Complex Organic Molecules (COMs) in Star-Forming Regions: A Virtual Special Issue}",
      journal = {ACS Earth and Space Chemistry},
         year = 2020,
        month = apr,
       volume = {4},
       number = {4},
        pages = {488-490},
          doi = {10.1021/acsearthspacechem.0c00043},
       adsurl = {https://ui.adsabs.harvard.edu/abs/2020ESC.....4..488.}
}

@ARTICLE{Jimenez-Serra22,
       author = {{Jim{\'e}nez-Serra}, Izaskun and {Rodr{\'\i}guez-Almeida}, Lucas F. and {Mart{\'\i}n-Pintado}, Jes{\'u}s and {Rivilla}, V{\'\i}ctor M. and {Melosso}, Mattia and {Zeng}, Shaoshan and {Colzi}, Laura and {Kawashima}, Yoshiyuki and {Hirota}, Eizi and {Puzzarini}, Cristina and {Tercero}, Bel{\'e}n and {de Vicente}, Pablo and {Rico-Villas}, Fernando and {Requena-Torres}, Miguel A. and {Mart{\'\i}n}, Sergio},
        title = "{Precursors of fatty alcohols in the ISM: Discovery of n-propanol}",
      journal = {\aap},
         year = 2022,
        month = jul,
       volume = {663},
          eid = {A181},
        pages = {A181},
          doi = {10.1051/0004-6361/202142699},
archivePrefix = {arXiv},
       eprint = {2204.08267},
 primaryClass = {astro-ph.GA},
       adsurl = {https://ui.adsabs.harvard.edu/abs/2022A&A...663A.181J}
}

@ARTICLE{Brown1988grains,
       author = {{Brown}, P.~D. and {Charnley}, S.~B. and {Millar}, T.~J.},
        title = "{A model of the chemistry in hot molecular cores.}",
      journal = {\mnras},
         year = 1988,
        month = mar,
       volume = {231},
        pages = {409-417},
          doi = {10.1093/mnras/231.2.409},
       adsurl = {https://ui.adsabs.harvard.edu/abs/1988MNRAS.231..409B}
}

@ARTICLE{Garrod2011,
       author = {{Garrod}, R.~T. and {Pauly}, T.},
        title = "{On the Formation of CO$_{2}$ and Other Interstellar Ices}",
      journal = {\apj},
         year = 2011,
        month = jul,
       volume = {735},
       number = {1},
          eid = {15},
        pages = {15},
          doi = {10.1088/0004-637X/735/1/15},
archivePrefix = {arXiv},
       eprint = {1106.0540},
 primaryClass = {astro-ph.GA},
       adsurl = {https://ui.adsabs.harvard.edu/abs/2011ApJ...735...15G}
}

@ARTICLE{Minissale2013,
       author = {{Minissale}, M. and {Congiu}, E. and {Manic{\`o}}, G. and {Pirronello}, V. and {Dulieu}, F.},
        title = "{CO$_{2}$ formation on interstellar dust grains: a detailed study of the barrier of the CO + O channel}",
      journal = {\aap},
         year = 2013,
        month = nov,
       volume = {559},
          eid = {A49},
        pages = {A49},
          doi = {10.1051/0004-6361/201321453},
       adsurl = {https://ui.adsabs.harvard.edu/abs/2013A&A...559A..49M}
}

@ARTICLE{Demaison2012,
       author = {{Demaison}, Jean and {Craig}, Norman C. and {Conrad}, Andrew R. and {Tubergen}, Michael J. and {Rudolph}, Heinz Dieter},
        title = "{Semiexperimental Equilibrium Structure of the Lower Energy Conformer of Glycidol by the Mixed Estimation Method}",
      journal = {Journal of Physical Chemistry A},
         year = 2012,
        month = sep,
       volume = {116},
       number = {36},
        pages = {9116-9122},
          doi = {10.1021/jp305504x},
       adsurl = {https://ui.adsabs.harvard.edu/abs/2012JPCA..116.9116D}
}

@ARTICLE{Jaman2015,
       author = {{Jaman}, A.~I. and {Chakraborty}, Shamik and {Chakraborty}, Rangana},
        title = "{Millimeterwave rotational spectrum and theoretical calculations of cis-propionic acid}",
      journal = {Journal of Molecular Structure},
         year = 2015,
        month = jan,
       volume = {1079},
        pages = {402-406},
          doi = {10.1016/j.molstruc.2014.09.004},
       adsurl = {https://ui.adsabs.harvard.edu/abs/2015JMoSt1079..402J}
}

@ARTICLE{Braakman2010,
       author = {{Braakman}, Rogier and {Drouin}, Brian J. and {Widicus Weaver}, Susanna L. and {Blake}, Geoffrey A.},
        title = "{Extended analysis of hydroxyacetone in the torsional ground state}",
      journal = {Journal of Molecular Spectroscopy},
         year = 2010,
        month = nov,
       volume = {264},
       number = {1},
        pages = {43-49},
          doi = {10.1016/j.jms.2010.09.003},
       adsurl = {https://ui.adsabs.harvard.edu/abs/2010JMoSp.264...43B}
}

@ARTICLE{Ishibashi2024,
       author = {{Ishibashi}, Atsuki and {Molpeceres}, Germ{\'a}n and {Hidaka}, Hiroshi and {Oba}, Yasuhiro and {Lamberts}, Thanja and {Watanabe}, Naoki},
        title = "{Proposed Importance of HOCO Chemistry: Inefficient Formation of CO$_{2}$ from CO and OH Reactions on Ice Dust}",
      journal = {\apj},
         year = 2024,
        month = dec,
       volume = {976},
       number = {2},
          eid = {162},
        pages = {162},
          doi = {10.3847/1538-4357/ad8235},
archivePrefix = {arXiv},
       eprint = {2410.01373},
 primaryClass = {astro-ph.GA},
       adsurl = {https://ui.adsabs.harvard.edu/abs/2024ApJ...976..162I}
}

@ARTICLE{Molpeceres2025,
       author = {{Molpeceres}, Germ{\'a}n and {Enrique-Romero}, Joan and {Ishibashi}, Atsuki and {Oba}, Yasuhiro and {Hidaka}, Hiroshi and {Lamberts}, Thanja and {Aikawa}, Yuri and {Watanabe}, Naoki},
        title = "{Hydrogenation of HOCO and formation of interstellar CO$_{2}$: a not so straightforward relation}",
      journal = {\mnras},
         year = 2025,
        month = apr,
       volume = {538},
       number = {3},
        pages = {1565-1575},
          doi = {10.1093/mnras/staf383},
archivePrefix = {arXiv},
       eprint = {2503.01692},
 primaryClass = {astro-ph.GA},
       adsurl = {https://ui.adsabs.harvard.edu/abs/2025MNRAS.538.1565M}
}

@ARTICLE{Jimenez-Serra25,
       author = {{Jim{\'e}nez-Serra}, Izaskun and {Meg{\'\i}as}, Andr{\'e}s and {Salaris}, Joseph and {Cuppen}, Herma and {Taillard}, Ang{\`e}le and {Jin}, Miwha and {Wakelam}, Valentine and {Vasyunin}, Anton I. and {Caselli}, Paola and {Pendleton}, Yvonne J. and {Dartois}, Emmanuel and {Noble}, Jennifer A. and {Viti}, Serena and {Borshcheva}, Katerina and {Garrod}, Robin T. and {Lamberts}, Thanja and {Fraser}, Helen and {Melnick}, Gary and {McClure}, Melissa and {Rocha}, Will and {Drozdovskaya}, Maria N. and {Lis}, Dariusz C.},
        title = "{Modelling methanol and hydride formation in the JWST Ice Age era}",
      journal = {\aap},
         year = 2025,
        month = mar,
       volume = {695},
          eid = {A247},
        pages = {A247},
          doi = {10.1051/0004-6361/202452389},
archivePrefix = {arXiv},
       eprint = {2502.10123},
 primaryClass = {astro-ph.GA},
       adsurl = {https://ui.adsabs.harvard.edu/abs/2025A&A...695A.247J}
}

@ARTICLE{Jimenez-Serra20,
       author = {{Jim{\'e}nez-Serra}, Izaskun and {Mart{\'\i}n-Pintado}, Jes{\'u}s and {Rivilla}, V{\'\i}ctor M. and {Rodr{\'\i}guez-Almeida}, Lucas and {Alonso Alonso}, Elena R. and {Zeng}, Shaoshan and {Cocinero}, Emilio J. and {Mart{\'\i}n}, Sergio and {Requena-Torres}, Miguel and {Mart{\'\i}n-Domenech}, Rafa and {Testi}, Leonardo},
        title = "{Toward the RNA-World in the Interstellar Medium{\textemdash}Detection of Urea and Search of 2-Amino-oxazole and Simple Sugars}",
      journal = {Astrobiology},
         year = 2020,
        month = sep,
       volume = {20},
       number = {9},
        pages = {1048-1066},
       adsurl = {https://ui.adsabs.harvard.edu/abs/2020AsBio..20.1048J}
}

@article{Medvedev:2009fs,
author = {Medvedev, Ivan R and De Lucia, Frank C and Herbst, Eric},
title = {{THE MILLIMETER- AND SUBMILLIMETER-WAVE SPECTRUM OF THE TRANSAND GAUCHECONFORMERS OF ETHYL FORMATE}},
journal = {Astrophysical Journal Supplement Series},
year = {2009},
volume = {181},
number = {2},
pages = {433--438},
month = mar
}

@article{Neill:2012fr,
author = {Neill, Justin L. and Muckle, Matt T. and Zaleski, Daniel P. and Steber, Amanda L. and Pate, Brooks H. and Lattanzi, Valerio and Spezzano, Silvia and McCarthy, Michael C. and Remijan, Anthony J.},
title = {{Laboratory and Tentative Interstellar Detection of Trans-Methyl Formate Using the Publicly Available Green Bank Telescope Primos Survey}},
journal = {\apj},
year = {2012},
volume = {755},
number = {2},
pages = {153},
month = aug
}

@article{Loomis:2015jh,
author = {Loomis, Ryan A and McGuire, Brett A and Shingledecker, Christopher and Johnson, Chelen H and Blair, Samantha and Robertson, Amy and Remijan, Anthony J.},
title = {{Investigating the minimum energy principle in searches for new molecular species{\textemdash}the case of H$_{2}$C$_{3}$O isomers}},
journal = {\apj},
year = {2015},
volume = {799},
number = {1},
pages = {34--8},
month = jan
}

@article{Demaison:1984if,
author = {Demaison, J and Boucher, D and BURIE, J and Dubrulle, A},
title = {{Millimeter-Wave Spectrum of Ethyl Formate}},
journal = {Zeitschrift f{\"u}r Naturforschung A},
year = {1984},
volume = {39},
number = {6},
pages = {560--564}
}

@article{Riveros:1967ic,
author = {Riveros, Jos{\'e} M},
title = {{Microwave Spectrum and Rotational Isomerism of Ethyl Formate}},
journal = {\jcp},
year = {1967},
volume = {46},
number = {12},
pages = {4605--9}
}

@article{Grimme2011,
author = {Grimme, Stefan and Ehrlich, Stephan and Goerigk, Lars},
title = {Effect of the damping function in dispersion corrected density functional theory},
journal = {Journal of Computational Chemistry},
volume = {32},
number = {7},
pages = {1456-1465},
year = {2011},
doi = {10.1002/jcc.21759},
url = {https://onlinelibrary.wiley.com/doi/abs/10.1002/jcc.21759},
eprint = {https://onlinelibrary.wiley.com/doi/pdf/10.1002/jcc.21759}
}

@article{Pickett1991,
title = {The fitting and prediction of vibration-rotation spectra with spin interactions},
journal = "J. Mol. Spectrosc. ",
volume = "148",
number = "2",
pages = "371 - 377",
year = "1991",
issn = "0022-2852",
author = "Herbert M. Pickett"
}

@ARTICLE{pickett1998,
       author = {{Pickett}, H.~M. and {Poynter}, R.~L. and {Cohen}, E.~A. and
         {Delitsky}, M.~L. and {Pearson}, J.~C. and {M{\"u}ller}, H.~S.~P.},
        title = "{Submillimeter, millimeter and microwave spectral line catalog.}",
      journal = {\jqsrt},
         year = 1998,
        month = nov,
       volume = {60},
       number = {5},
        pages = {883-890},
       adsurl = {https://ui.adsabs.harvard.edu/abs/1998JQSRT..60..883P}
}

@ARTICLE{sanz-novo2022,
	author = {{Sanz-Novo, M.} and {Belloche, A.} and {Rivilla, V. M.} and {Garrod, R. T.} and {Alonso, J. L.} and {Redondo, P.} and {Barrientos, C.} and {Kolesniková, L.} and {Valle, J. C.} and {Rodríguez-Almeida, L.} and {Jimenez-Serra, I.} and {Martín-Pintado, J.} and {Müller, H. S. P.} and {Menten, K. M.}},
	title = {Toward the limits of complexity of interstellar chemistry: Rotational spectroscopy and astronomical search for n- and i-butanal★},
	DOI= "10.1051/0004-6361/202142848",
	url= "https://doi.org/10.1051/0004-6361/202142848",
	journal = {{\aap}},
	year = 2022,
	volume = 666,
	pages = "A114",
}

@article{endres_cologne_2016,
	title = {The {Cologne} {Database} for {Molecular} {Spectroscopy}, {CDMS}, in the {Virtual} {Atomic} and {Molecular} {Data} {Centre}, {VAMDC}},
	volume = {327},
	issn = {0022-2852},
	url = {http://adsabs.harvard.edu/abs/2016JMoSp.327...95E},
	doi = {10.1016/j.jms.2016.03.005},
	journal = {Journal of Molecular Spectroscopy},
	author = {Endres, Christian P. and Schlemmer, Stephan and Schilke, Peter and Stutzki, Jürgen and Müller, Holger S. P.},
	month = sep,
	year = {2016},
	pages = {95--104}
}

@article{rivilla_chemical_2017,
	title = {On the chemical ladder of esters. {Detection} and formation of ethyl formate in the {W}51 e2 hot molecular core},
	volume = {599},
	issn = {0004-6361},
	url = {http://adsabs.harvard.edu/abs/2017A%26A...599A..26R},
	journal = {\aap},
	author = {Rivilla, V. M. and Beltr\'an, M. T. and Mart\'in-Pintado, J. and Fontani, F. and Caselli, P. and Cesaroni, R.},
	month = mar,
	year = {2017},
	pages = {A26}
}

@article{martin_tracing_2008,
	title = {Tracing shocks and photodissociation in the {Galactic} center region},
	volume = {678},
	issn = {0004-637X, 1538-4357},
	number = {1},
	journal = {\apj},
	author = {Mart\'in, S. and Requena-Torres, M. A. and Mart\'in-Pintado, J. and Mauersberger, R.},
	month = may,
	year = {2008},
	pages = {245--254},
}

@article{requena-torres_largest_2008,
	title = {The largest oxigen bearing organic molecule repository},
	volume = {672},
	issn = {0004-637X, 1538-4357},
	number = {1},
	journal = {\apj},
	author = {Requena-Torres, M. A. and Mart\'in-Pintado, J. and Mart\'in, S. and Morris, M. R.},
	month = jan,
	year = {2008},
	pages = {352--360},
}

@article{requena-torres_organic_2006,
	title = {Organic {Molecules} in the {Galactic} {Center}. {Hot} {Core} {Chemistry} without {Hot} {Cores}},
	volume = {455},
	issn = {0004-6361, 1432-0746},
	url = {http://arxiv.org/abs/astro-ph/0605031},
	doi = {10.1051/0004-6361:20065190},
	number = {3},
	journal = {A\&A},
	author = {Requena-Torres, M. A. and Mart\'in-Pintado, J. and Rodr\'iguez-Franco, A. and Mart\'in, S. and Rodr\'iguez-Fern\'andez, N. J. and de Vicente, P.},
	month = sep,
	year = {2006},
	pages = {971--985},
}

@ARTICLE{shingledecker2019,
       author = {{Shingledecker}, Christopher N. and {{\'A}lvarez-Barcia}, Sonia and
         {Korn}, Viktoria H. and {K{\"a}stner}, Johannes},
        title = "{The Case of H$_{2}$C$_{3}$O Isomers, Revisited: Solving the Mystery of the Missing Propadienone}",
      journal = {\apj},
         year = "2019",
        month = "Jun",
       volume = {878},
       number = {2},
          eid = {80},
        pages = {80},
          doi = {10.3847/1538-4357/ab1d4a},
archivePrefix = {arXiv},
       eprint = {1904.11396},
 primaryClass = {astro-ph.GA},
       adsurl = {https://ui.adsabs.harvard.edu/abs/2019ApJ...878...80S}
}

@ARTICLE{martin2019,
       author = {{Mart{\'\i}n}, S. and {Mart{\'\i}n-Pintado}, J. and
         {Blanco-S{\'a}nchez}, C. and {Rivilla}, V.~M. and
         {Rodr{\'\i}guez-Franco}, A. and {Rico-Villas}, F.},
        title = "{Spectral Line Identification and Modelling (SLIM) in the MAdrid Data CUBe Analysis (MADCUBA) package. Interactive software for data cube analysis}",
      journal = {\aap},
         year = "2019",
        month = "Nov",
       volume = {631},
          eid = {A159},
        pages = {A159},
       adsurl = {https://ui.adsabs.harvard.edu/abs/2019A&A...631A.159M}
}

@ARTICLE{zeng2018,
   author = {{Zeng}, S. and {Jim{\'e}nez-Serra}, I. and {Rivilla}, V.~M. and 
	{Mart{\'{\i}}n}, S. and {Mart{\'{\i}}n-Pintado}, J. and {Requena-Torres}, M.~A. and 
	{Armijos-Abenda{\~n}o}, J. and {Riquelme}, D. and {Aladro}, R.
	},
    title = "{Complex organic molecules in the Galactic Centre: the N-bearing family}",
  journal = {\mnras},
     year = 2018,
    month = aug,
   volume = 478,
    pages = {2962-2975},
   adsurl = {http://adsabs.harvard.edu/abs/2018MNRAS.478.2962Z}
}

@ARTICLE{rodriguez-almeida2021b,
       author = {{Rodr{\'\i}guez-Almeida}, L.~F. and {Rivilla}, V.~M. and {Jim{\'e}nez-Serra}, I. and {Melosso}, M. and {Colzi}, L. and {Zeng}, S. and {Tercero}, B. and {de Vicente}, P. and {Mart{\'\i}n}, S. and {Requena-Torres}, M.~A. and {Rico-Villas}, F. and {Mart{\'\i}n-Pintado}, J.},
        title = "{First detection of C$_{2}$H$_{5}$NCO in the ISM and search of other isocyanates towards the G+0.693-0.027 molecular cloud}",
      journal = {\aap},
         year = 2021,
        month = oct,
       volume = {654},
          eid = {L1},
        pages = {L1},
       adsurl = {https://ui.adsabs.harvard.edu/abs/2021A&A...654L...1R}
}

@ARTICLE{rodriguez-almeida2021a,
       author = {{Rodr{\'\i}guez-Almeida}, Lucas F. and {Jim{\'e}nez-Serra}, Izaskun and {Rivilla}, V{\'\i}ctor M. and {Mart{\'\i}n-Pintado}, Jes{\'u}s and {Zeng}, Shaoshan and {Tercero}, Bel{\'e}n and {de Vicente}, Pablo and {Colzi}, Laura and {Rico-Villas}, Fernando and {Mart{\'\i}n}, Sergio and {Requena-Torres}, Miguel A.},
        title = "{Thiols in the Interstellar Medium: First Detection of HC(O)SH and Confirmation of C$_{2}$H$_{5}$SH}",
      journal = {\apjl},
         year = 2021,
        month = may,
       volume = {912},
       number = {1},
          eid = {L11},
        pages = {L11},
       adsurl = {https://ui.adsabs.harvard.edu/abs/2021ApJ...912L..11R}
}

@ARTICLE{zeng2020,
    author = {Zeng, S and Zhang, Q and Jiménez-Serra, I and Tercero, B and Lu, X and Martín-Pintado, J and de Vicente, P and Rivilla, V M and Li, S},
    title = "{Cloud–cloud collision as drivers of the chemical complexity in Galactic Centre molecular clouds}",
    journal = {\mnras},
    volume = {497},
    number = {4},
    pages = {4896-4909},
    year = {2020},
    month = {07},
    issn = {0035-8711},
    doi = {10.1093/mnras/staa2187},
    url = {https://doi.org/10.1093/mnras/staa2187},
    eprint = {https://academic.oup.com/mnras/article-pdf/497/4/4896/33684472/staa2187.pdf},
}

@ARTICLE{tercero2021,
       author = {{Tercero}, F. and {L{\'o}pez-P{\'e}rez}, J.~A. and {Gallego}, J.~D. and {Beltr{\'a}n}, F. and {Garc{\'\i}a}, O. and {Patino-Esteban}, M. and {L{\'o}pez-Fern{\'a}ndez}, I. and {G{\'o}mez-Molina}, G. and {Diez}, M. and {Garc{\'\i}a-Carre{\~n}o}, P. and {Malo}, I. and {Amils}, R. and {Serna}, J.~M. and {Albo}, C. and {Hern{\'a}ndez}, J.~M. and {Vaquero}, B. and {Gonz{\'a}lez-Garc{\'\i}a}, J. and {Barbas}, L. and {L{\'o}pez-Fern{\'a}ndez}, J.~A. and {Bujarrabal}, V. and {G{\'o}mez-Garrido}, M. and {Pardo}, J.~R. and {Santander-Garc{\'\i}a}, M. and {Tercero}, B. and {Cernicharo}, J. and {de Vicente}, P.},
        title = "{Yebes 40 m radio telescope and the broad band Nanocosmos receivers at 7 mm and 3 mm for line surveys}",
      journal = {\aap},
         year = 2021,
        month = jan,
       volume = {645},
          eid = {A37},
        pages = {A37},
          doi = {10.1051/0004-6361/202038701},
archivePrefix = {arXiv},
       eprint = {2010.16224},
 primaryClass = {astro-ph.IM},
       adsurl = {https://ui.adsabs.harvard.edu/abs/2021A&A...645A..37T}
}

@ARTICLE{Rivilla2022c,
  
AUTHOR={Rivilla, Víctor M. and Jiménez-Serra, Izaskun and Martín-Pintado, Jesús and Colzi, Laura and Tercero, Belén and de Vicente, Pablo and Zeng, Shaoshan and Martín, Sergio and García de la Concepción, Juan and Bizzocchi, Luca and Melosso, Mattia and Rico-Villas, Fernando and Requena-Torres, Miguel A.},   
	 
TITLE={Molecular Precursors of the RNA-World in Space: New Nitriles in the G+0.693−0.027 Molecular Cloud},      
	
JOURNAL={Frontiers in Astronomy and Space Sciences},      
	
VOLUME={9:829288},           
	
YEAR={2022c},      
	  
URL={https://www.frontiersin.org/articles/10.3389/fspas.2022.876870},       
	
DOI={10.3389/fspas.2022.876870},      
	
ISSN={2296-987X}
}

@ARTICLE{rivilla2022a,
       author = {{Rivilla}, V{\'\i}ctor M. and {Colzi}, Laura and {Jim{\'e}nez-Serra}, Izaskun and {Mart{\'\i}n-Pintado}, Jes{\'u}s and {Meg{\'\i}as}, Andr{\'e}s and {Melosso}, Mattia and {Bizzocchi}, Luca and {L{\'o}pez-Gallifa}, {\'A}lvaro and {Mart{\'\i}nez-Henares}, Antonio and {Massalkhi}, Sarah and {Tercero}, Bel{\'e}n and {de Vicente}, Pablo and {Guillemin}, Jean-Claude and {Garc{\'\i}a de la Concepci{\'o}n}, Juan and {Rico-Villas}, Fernando and {Zeng}, Shaoshan and {Mart{\'\i}n}, Sergio and {Requena-Torres}, Miguel A. and {Tonolo}, Francesca and {Alessandrini}, Silvia and {Dore}, Luca and {Barone}, Vincenzo and {Puzzarini}, Cristina},
        title = "{Precursors of the RNA World in Space: Detection of (Z)-1,2-ethenediol in the Interstellar Medium, a Key Intermediate in Sugar Formation}",
      journal = {\apjl},
         year = 2022,
        month = apr,
       volume = {929},
       number = {1},
          eid = {L11},
        pages = {L11},
       adsurl = {https://ui.adsabs.harvard.edu/abs/2022ApJ...929L..11R}
}

@ARTICLE{mininni2020,
       author = {{Mininni}, C. and {Beltr{\'a}n}, M.~T. and {Rivilla}, V.~M. and {S{\'a}nchez-Monge}, A. and {Fontani}, F. and {M{\"o}ller}, T. and {Cesaroni}, R. and {Schilke}, P. and {Viti}, S. and {Jim{\'e}nez-Serra}, I. and {Colzi}, L. and {Lorenzani}, A. and {Testi}, L.},
        title = "{The GUAPOS project: G31.41+0.31 Unbiased ALMA sPectral Observational Survey. I. Isomers of C$_{2}$H$_{4}$O$_{2}$}",
      journal = {\aap},
         year = 2020,
        month = dec,
       volume = {644},
          eid = {A84},
        pages = {A84},
          doi = {10.1051/0004-6361/202038966},
archivePrefix = {arXiv},
       eprint = {2009.13297},
 primaryClass = {astro-ph.SR},
       adsurl = {https://ui.adsabs.harvard.edu/abs/2020A&A...644A..84M}
}
\bibliographystyle{aa}

\newpage

\begin{appendix}
 
\section{Complementary Tables and Figures}
\label{rot_backgr}

In Tables \ref{tab:LTEMA}, \ref{tab:aethylformate}, \ref{tab:gethylformate}, \ref{tab:hydroxyacetone}, \ref{tab:lact} and \ref{tab:meta} we provide the spectroscopic information of the transitions detected for methyl acetate, $a$-ethyl formate, $g$-ethyl formate, hydroxyacetone, lactaldehyde and methoxyacetaldehyde, respectively, which are shown in Figures \ref{f:LTEMA}, \ref{f:LTEaEF}, \ref{f:LTEgEF}, \ref{f:LTEHA}, \ref{f:LTELA} and \ref{f:LTEMeta}.

Regarding the preparation of the line catalog of methyl acetate in SPCAT format from the BELGI output, we also provide in Table \ref{tab:LTEMA} the values of the line strength ($S_g \mu_g^2$; \citealt{methyl_acetate_Orion}), which were used to compute the log \textit{I}(nm$^2$ MHz) at 300 K via:

{\small
\[
I(T) = 4.16231 \cdot 10^{-5} \cdot \nu \cdot g_I \cdot S_g \cdot \mu_g^2 \cdot \frac{\exp(-E''/kT) - \exp(-E'/kT)}{Q_{\text{r}}(T)}
\]
}

with $\nu$ in MHz and $\mu$ in D. 

We note that the original laboratory data covered up to $J$ = 19 and $K$$_a$ $\leq$ 7, although predictions were initially computed up to $J$ = 30 \citep{methyl_acetate_Orion}. In this work, we adopt the rotational partition function values computed from first principles, derived from the rotational constants reported in \cite{Nguyen2014}. Specifically, we obtained a $Q$$_r$ (AA, 300 K) = 72174.2179 (without considering spin weights and all states), which is approximately 15$\%$ larger than the value used in \cite{methyl_acetate_Orion} of $Q$$_r$ (AA, 300 K) = 61648.6 (I. Kleiner, priv. communication). The latter was obtained by applying a correction factor of 2.6 to the $Q$$_r$ (AA) computed through direct summation of the ground-state energy levels up to $J$ = 30, which is needed to consider energy levels up to $J$ = 65. Nevertheless, it may still underestimate the real partition function value if full convergence has not been achieved \citep{Carvajal2019}. 

In contrast to the previous work, where each substate was treated as a separate species \citep{methyl_acetate_Orion}, we have included all five substates in the same catalog. Consequently, the $E$$\mathrm{_{up}}$ (K) reported in Table \ref{tab:LTEMA} accounts for the ZPE of the AA species (99.9450 cm$^{-1}$; \citealt{methyl_acetate_Orion}). To consider all of the states and their weights, $Q$$_r$(Tot, 300 K) = $Q$$_r$ (AA, 300 K) $\times$ f, where f = (16+16+16+8+8) = 64.

In Figure \label{f:LTEgEF} we show the tentative detection of \textit{g}-ethyl formate, $g$-\ch{CH3CH2OC(O)H}, toward G+0.693, while in Figures \ref{f:nondetectionPA} and \ref{f:nondetectionGly} we present the LTE simulation of the propionic acid and glycidol emission, respectively, using the upper limits to the column density derived toward G+0.693. 


\begin{center}
\begin{figure*}[ht]
     \centerline{\resizebox{0.8
     \hsize}{!}{\includegraphics[angle=0]{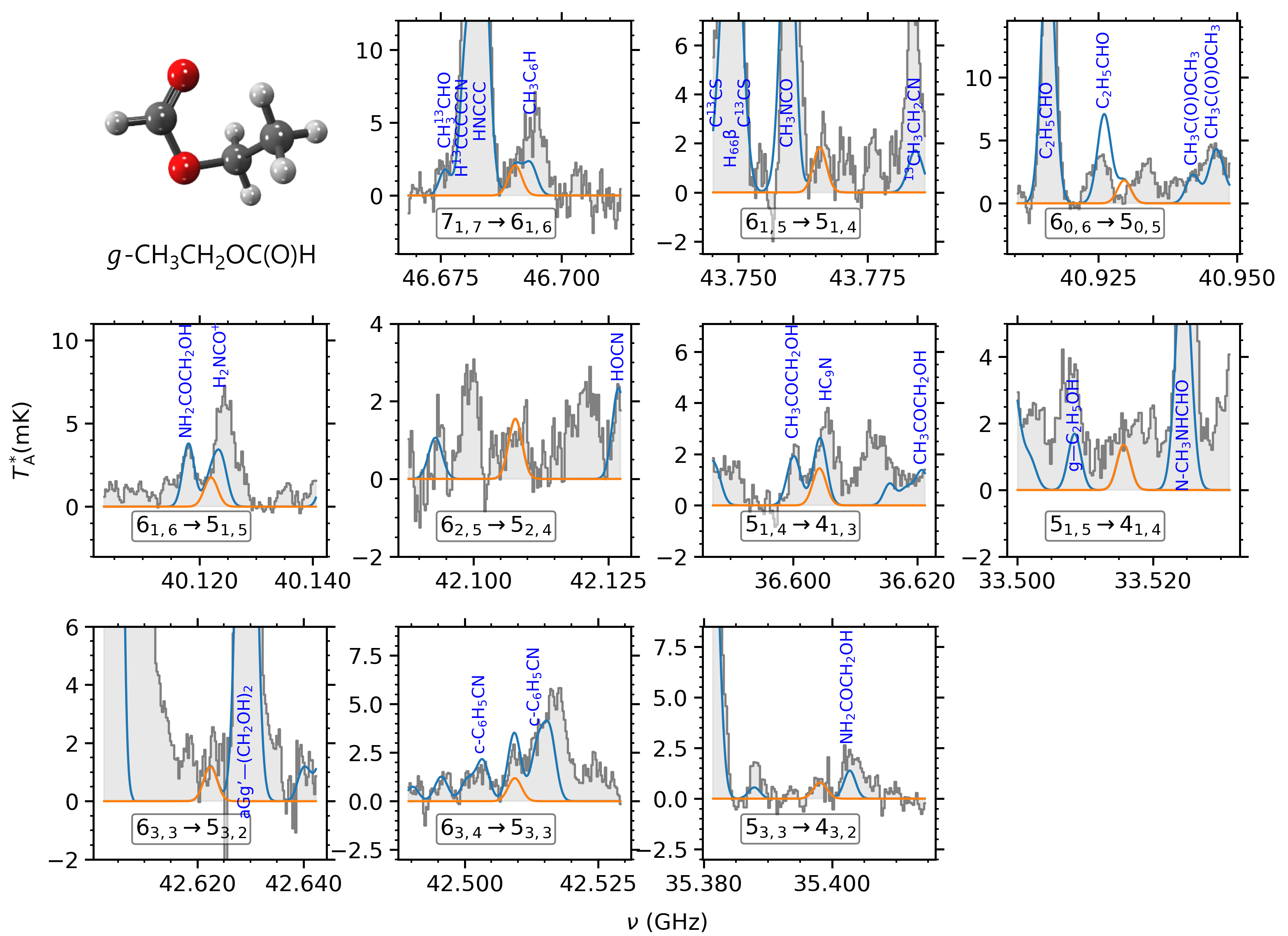}}}
     \caption{Tentative detection of \textit{g}-ethyl formate, $g$-\ch{CH3CH2OC(O)H}, toward G+0.693. The selected transitions are sorted by decreasing intensity and listed in Table \ref{tab:gethylformate}. The result of the LTE model of the $g$-\ch{CH3CH2OC(O)H} emission is shown in orange, together with the expected molecular emission from all the molecular species identified to date in our survey, including  $g$-\ch{CH3CH2OC(O)H} (in blue), both overlaid on the observations (gray histogram). The molecular structure of $g$-\ch{CH3CH2OC(O)H} is also shown (C atoms in gray, O atoms in red and H atoms in white).}
\label{f:LTEgEF}
\end{figure*}
\end{center}

\begin{figure*}
\centerline{\resizebox{0.7\hsize}{!}{\includegraphics[angle=0]{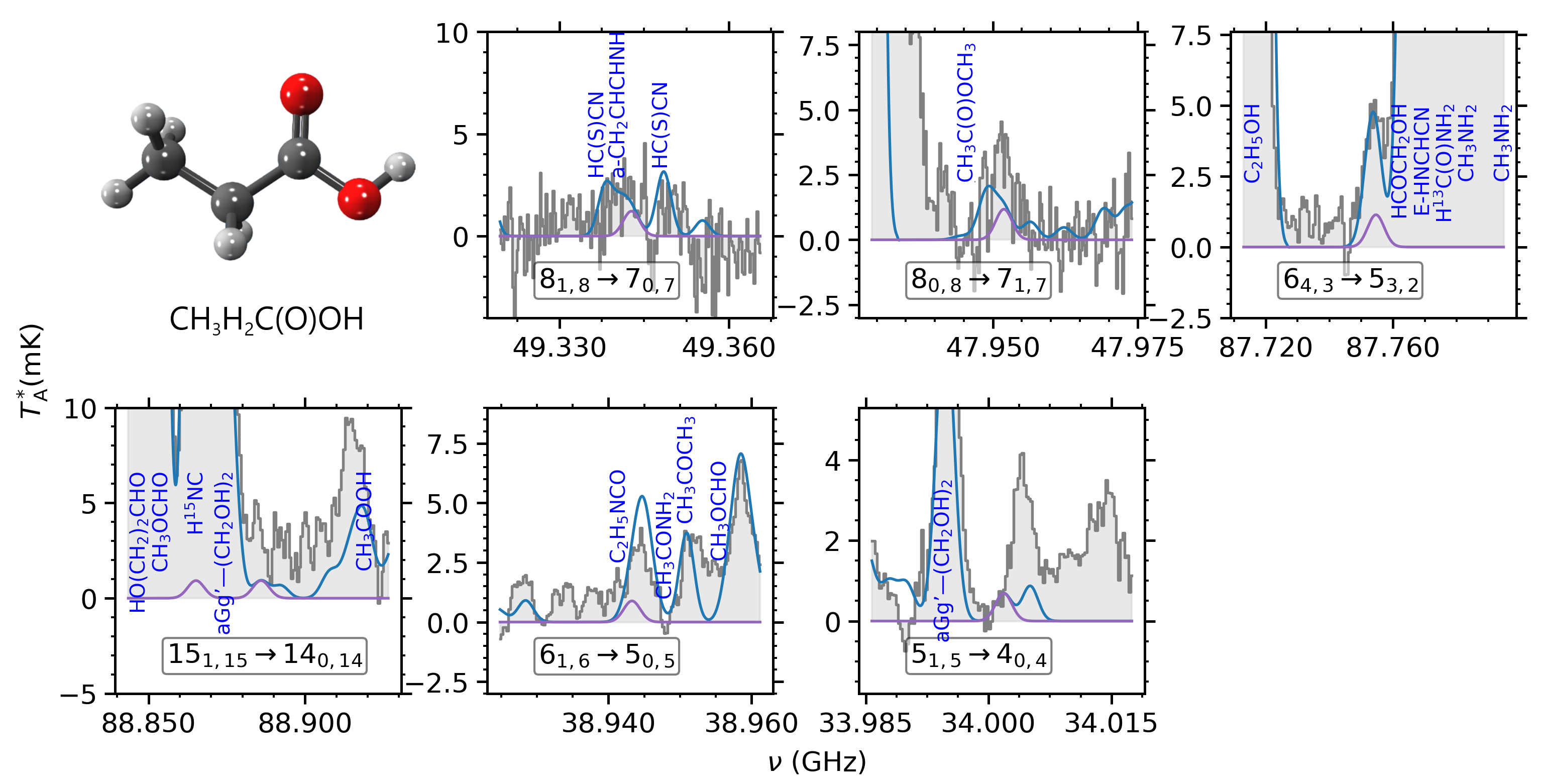}}}
\caption{LTE simulation of the propionic acid, \ch{CH3CH2C(O)OH}, emission using the upper limit column density derived toward G+0.693 (in purple) together with the expected molecular emission from all the molecular species identified to date in our survey (in blue), both overlaid on the observations (gray histogram). The quantum numbers for each transition are shown at the bottom of each panel.}
\label{f:nondetectionPA}
\end{figure*}

\begin{figure*}
\centerline{\resizebox{0.7\hsize}{!}{\includegraphics[angle=0]{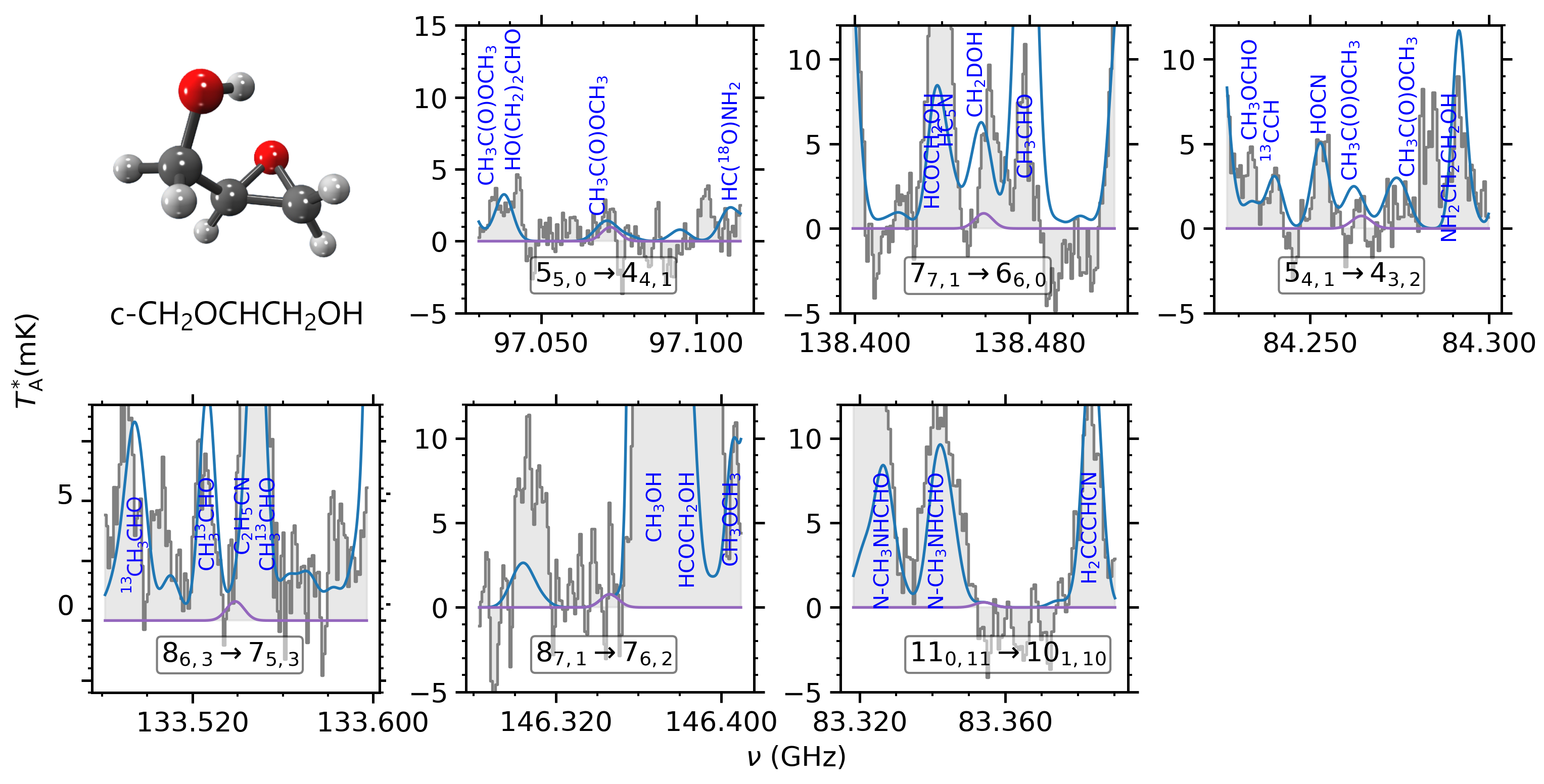}}}
\caption{LTE simulation of the glycidol, (c-\ch{CH2OCHCH2OH}, emission using the upper limit column density derived toward G+0.693 (in purple) together with the expected molecular emission from all the molecular species identified to date in our survey (in blue), both overlaid on the observations (gray histogram). The quantum numbers for each transition are shown at the bottom of each panel.}
\label{f:nondetectionGly}
\end{figure*}

\clearpage

\begin{table*}
\tabcolsep 4pt
\centering
\caption{Spectroscopic information of the selected transitions of methyl acetate detected toward G+0.693, shown in Figure \ref{f:LTEMA}.} 
\label{tab:LTEMA}
\begin{tabular}{cccccccc}
\hline 
Frequency (GHz)  & Transition$^{(a)}$ & $S_g \mu_g^2$ & log \textit{I} at 300 K (nm$^2$ MHz)& $E$$\mathrm{_{up}}$ (K) & Blending \\ 
\hline
32.797604 (2)  &  6$_{3, 3}$--6$_{2, 4}$ (E$_3$)  &  8.19 &  --7.0096 & 12.2 & Unblended \\     
32.840814 (2)  &  6$_{3, 3}$--6$_{2, 4}$ (E$_1$)  &  8.20 & --7.0081  & 12.2 & Slightly blended: U-line \\ 
34.628602 (2)  &  4$_{3, 1}$--4$_{2, 2}$ (E$_3$)  &  4.70 & --7.1979  & 8.3 & Unblended\\     
34.637085 (4)  &  8$_{3, 6}$--8$_{2, 7}$ (E$_4$)  &  9.69 & --6.8957  & 16.9 & Slightly blended: U-line \\ 
34.715239 (4)  &  8$_{3, 6}$--8$_{2, 7}$ (E$_3$)  &  9.67 & --6.8947  & 16.9 & Unblended \\     
35.446462 (2)  &  5$_{1, 5}$--4$_{0, 4}$ (E$_4$)  &  9.56 & --6.8672  & 6.7 & Slightly blended: U-line \\  
35.452045 (2)  &  5$_{1, 5}$--4$_{0, 4}$ (E$_1$)  &  9.56 & --6.8669  & 6.7 & Unblended \\     
35.454312 (2)  &  5$_{1, 5}$--4$_{0, 4}$ (E$_3$)  &  9.56 & --6.8667  & 6.7 & Unblended \\     
38.523669 (4)  &  9$_{3, 7}$--9$_{2, 8}$ (E$_1$)  & 10.81 & --6.7608  & 20.1 & Blended: U line \\     
38.537458 (2)  &  6$_{0, 6}$--5$_{1, 5}$ (E$_3$)  & 11.85 & --6.7036  & 8.6 & Unblended \\     
38.539998 (2)  &  6$_{0, 6}$--5$_{1, 5}$ (E$_1$)  & 11.85 & --6.7036  & 8.6 & Unblended\\     
38.543524 (2)  &  6$_{0, 6}$--5$_{1, 5}$ (E$_4$)  & 11.85 & --6.7036  & 8.6 & Slightly blended: U-line \\ 
38.549823 (4)  &  9$_{3, 7}$--9$_{2, 8}$ (E$_3$)  & 10.80 & --6.7603  & 20.1 & Blended: U-line \\     
40.941899 (2)  &  6$_{1, 6}$--5$_{0, 5}$ (E$_4$)  & 12.20 & --6.6386  & 8.6 & Slightly blended: U-line \\ 
40.945609 (2)  &  6$_{1, 6}$--5$_{0, 5}$ (E$_1$)  & 12.20 & --6.6385  & 8.6 & Unblended  \\     
40.946537 (2)  &  6$_{1, 6}$--5$_{0, 5}$ (E$_3$)  & 12.20 & --6.6384  & 8.6 & Unblended\\     
41.235605 (2)  &  6$_{1, 6}$--5$_{0, 5}$ (E$_2$)  & 12.29 & --6.6266  & 7.0 & Unblended \\     
41.237259 (2)  &  6$_{1, 6}$--5$_{0, 5}$ (AA)     & 12.29 & --6.6265  & 7.0 & Unblended \\     
42.640369 (4)  &  6$_{4, 3}$--6$_{3, 4}$ (E$_4$)  & 7.13  & --6.8433  & 13.6 & Slightly blended: U-line \\   
46.805363 (2)  &  7$_{1, 7}$--6$_{0, 6}$ (E$_2$)  & 14.97 & --6.4339  & 9.2 & Unblended \\     
46.805610 (10) & 12$_{2,10}$--12$_{1,11}$ (E$_3$) & 13.55 & --6.5093 & 31.4 & Unblended \\     
46.806655 (2)  &  7$_{1, 7}$--6$_{0, 6}$ (AA)     & 14.97 & --6.4338 & 9.2 & Unblended  \\     
46.811991 (10) & 12$_{2,10}$--12$_{1,11}$ (E$_4$) & 13.55 & --6.5093 & 31.4 & Slightly blended: U-line \\  
46.812678 (10) & 12$_{2,10}$--12$_{1,11}$ (E$_1$) & 13.55 & --6.5091 & 31.4 & Slightly blended: U-line \\
76.613177 (9)  & 12$_{0,12}$--12$_{1,11}$ (E$_3$) & 28.59 & --5.7486 & 26.3 & Slightly blended: NH$_2$CH$_2$CH$_2$OH\\  
76.613332 (9)  & 12$_{0,12}$--12$_{1,11}$ (E$_4$) & 28.59 & --5.7486 & 26.3 & Slightly blended: NH$_2$CH$_2$CH$_2$OH\\  
76.613594 (9)  & 12$_{0,12}$--12$_{1,11}$ (E$_1$) & 28.59 & --5.7485 & 26.2 & Slightly blended: NH$_2$CH$_2$CH$_2$OH\\  
83.789892 (4)  &  5$_{4, 1}$--4$_{3, 2}$ (AA)     & 9.26  & --6.1373 & 10.4 & Slightly blended: C$_2$H$_5$OCHO\\ 
88.939251 (15) & 14$_{0,14}$--13$_{1,13}$ (E$_3$) & 34.01 & --5.5552 & 34.5 & Blended: t-C$_2$H$_3$CHO\\   
88.939307 (15) & 14$_{0,14}$--13$_{1,13}$ (E$_4$) & 34.01 & --5.5552 & 34.5 & Blended: t-C$_2$H$_3$CHO \\  
88.939623 (15) & 14$_{0,14}$--13$_{1,13}$ (E$_1$) & 34.01 & --5.5551 & 34.5 & Blended: t-C$_2$H$_3$CHO\\   
88.953729 (15) & 14$_{1,14}$--13$_{0,13}$ (E$_3$) & 34.01 & --5.5550 & 34.5 & Unblended \\   
88.953732 (15) & 14$_{1,14}$--13$_{0,13}$ (E$_4$) & 34.01 & --5.5550 & 34.5 & Unblended \\ 
88.954100 (15) & 14$_{1,14}$--13$_{0,13}$ (E$_1$) & 34.01 & --5.5550 & 34.5 & Unblended \\    
94.160620 (5)  &  7$_{4, 4}$--6$_{3, 4}$ (E$_1$)  & 9.13  & --6.0496  & 16.0 & Unblended\\      
94.166317 (5)  &  7$_{4, 4}$--6$_{3, 4}$ (E$_3$)  & 9.13  & --6.0495  & 16.1 & Unblended\\    
94.313817 (14) & 14$_{1,13}$--13$_{2,12}$ (E$_3$) & 26.70 & --5.6141  & 37.9 & Slightly blended: CH$_3$$^{13}$CH$_2$OH\\  
94.314988 (14) & 14$_{1,13}$--13$_{2,12}$ (E$_4$) & 26.70 & --5.6141  & 37.9 & Slightly blended: CH$_3$$^{13}$CH$_2$OH \\ 
94.315443 (14) & 14$_{1,13}$--13$_{2,12}$ (E$_1$) & 26.70 & --5.6141  & 37.9 & Slightly blended: CH$_3$$^{13}$CH$_2$OH \\
95.078847 (16) & 14$_{2,13}$--13$_{1,12}$ (E$_2$) & 26.65 & --5.6057  & 36.4 & Slightly blended: HCOCH$_2$OH\\  
95.080895 (16) & 14$_{2,13}$--13$_{1,12}$ (AA)    & 26.65 & --5.6057  & 36.4 & Unblended\\      
95.095381 (18) & 15$_{0,15}$--14$_{1,14}$ (E$_3$) & 36.71 & --5.4702  & 39.1 & Unblended\\      
95.095417 (18) & 15$_{0,15}$--14$_{1,14}$ (E$_4$) & 36.71 & --5.4702  & 39.1 & Unblended\\      
95.095735 (18) & 15$_{0,15}$--14$_{1,14}$ (E$_1$) & 36.71 & --5.4702  & 39.1 & Unblended\\           
95.102484 (18) & 15$_{1,15}$--14$_{0,14}$ (E$_3$) & 36.71 & --5.4701  & 39.1 & Unblended\\      
95.102496 (18) & 15$_{1,15}$--14$_{0,14}$ (E$_4$) & 36.71 & --5.4702  & 39.1 & Unblended\\      
95.102840 (18) & 15$_{1,15}$--14$_{0,14}$ (E$_1$) & 36.71 & --5.4701  & 39.1 & Unblended\\      
98.823558 (5)  &  7$_{4, 3}$--6$_{3, 3}$ (E$_1$)  & 9.40  & --5.9965  & 17.0 & Slightly blended: CH$_3$COCH$_3$ \\   
98.826693 (4)  &  7$_{4, 3}$--6$_{3, 3}$ (E$_4$)  & 9.40  & --5.9965  & 17.0 & Slightly blended: CH$_3$COCH$_3$ \\ 
101.249746 (22)& 16$_{0,16}$--15$_{1,15}$ (E$_3$) & 39.41 & --5.3917  & 43.9 &  Slightly blended: C$_2$H$_5$SH\\      
101.249772 (22)& 16$_{0,16}$--15$_{1,15}$ (E$_4$) & 39.41 & --5.3917  & 43.9 &  Slightly blended: C$_2$H$_5$SH\\      
101.250085 (22)& 16$_{0,16}$--15$_{1,15}$ (E$_1$) & 39.41 & --5.3917  & 43.9 &  Slightly blended: C$_2$H$_5$SH\\      
101.253205 (22)& 16$_{1,16}$--15$_{0,15}$ (E$_3$) & 39.41 & --5.3917  & 43.9 &  Slightly blended: C$_2$H$_5$SH\\      
101.253219 (22)& 16$_{1,16}$--15$_{0,15}$ (E$_4$) & 39.41 & --5.3917  & 43.9 &  Slightly blended: C$_2$H$_5$SH\\ 
101.253545 (22)& 16$_{1,16}$--15$_{0,15}$ (E$_1$) & 39.41 & --5.3917  & 43.9 &  Slightly blended: C$_2$H$_5$SH\\  
\hline
\end{tabular}
\vspace*{-0.5ex}
\tablefoot{$^{(a)}$ The AA, E$_1$, E$_2$ and E$_3$ and E$_4$ labels refer to the different symmetry or torsional substates, arising from the presence of two non-equivalent CH$_3$ internal rotation motions. Numbers in parentheses represent the predicted uncertainty associated to the last digits. We denote as U-line the line blending with a yet unidentified feature, and as “unblended" lines those that are not contaminated by other species, although many transitions belonging to different symmetry substates are partially or fully coalesced (auto-blended).}\\
\end{table*}

\begin{table*}
\tabcolsep 5pt
\centering
\caption{Spectroscopic information of the transitions of $a$-ethyl formate detected toward G+0.693$-$0.027, shown in Figure \ref{f:LTEaEF}.}
\begin{tabular}{cccccccccccc}
\hline
Frequency & Transition$^{(a)}$ & log \textit{I} (300 K) & $E$$\mathrm{_{up}}$  & Blending  \\ 
(GHz) &                        &  (nm$^2$ MHz)          &  (K)                 &    \\
\hline
31.8852150 (19) & 6$_{1,6}$--5$_{1,5}$  & -6.1602 &  6.0 & Blended: C$_6$H  \\ 
32.7188633 (19) & 6$_{0,6}$--5$_{0,5}$  & -6.1251 &  5.5 & Blended: H$_{83}$$\gamma$ \\ 
32.8790743 (19) & 6$_{2,5}$--5$_{2,4}$  & -6.1757 &  8.3 & Unblended \\ 
33.8355565 (19) & 6$_{1,5}$--5$_{1,4}$  & -6.1090 &  6.4 & Slightly blended: CH$_3$CHCO \\ 
38.3453139 (22) & 7$_{2,6}$--6$_{2,5}$  & -5.9635 & 10.2 & Slightly blended: HCCCH$_2$CN \\ 
38.4274410 (22) & 7$_{3,5}$--6$_{3,4}$  & -6.0180 & 13.8 & Blended: HCCCH$_2$CN \\   
38.4342157 (22) & 7$_{3,4}$--6$_{3,3}$  & -6.0178 & 13.8 & Slightly blended: U-line \\ 
38.6374107 (22) & 7$_{2,5}$--6$_{2,4}$  & -5.9570 & 10.2 & Blended: H$_{86}$$\delta$ \\ 
39.4528074 (22) & 7$_{1,6}$--6$_{1,5}$  & -5.9080 &  8.2 & Slightly blended: U-line \\ 
42.4683643 (25) & 8$_{1,8}$--7$_{1,7}$  & -5.7861 &  9.8 & Blended: $^{13}$CH$_3$CH$_2$CN \\ 
43.8052932 (25) & 8$_{2,7}$--7$_{2,6}$  & -5.7838 & 12.3 & Slightly blended: HCCCH$_2$CN \\ 
44.2387698 (25) & 8$_{2,6}$--7$_{2,5}$  & -5.7753 & 12.3 & Slightly blended: U-line *\\ 
45.0591299 (25) & 8$_{1,7}$--7$_{1,6}$  & -5.7354 & 10.4 & Unblended \\ 
48.7428006 (27) & 9$_{0,9}$--8$_{0,8}$  & -5.6117 & 11.7 & Unblended \\ 
49.2581397 (27) & 9$_{2,8}$--8$_{2,7}$  & -5.6280 & 14.6 & Slightly blended: HCCCH$_2$CN \\ 
49.8680897 (27) & 9$_{2,7}$--8$_{2,6}$  & -5.6174 & 14.7 & Unblended \\ 
\hline 
\end{tabular}
\label{tab:aethylformate}
\vspace*{-0.5ex}
\tablefoot{$^{(a)}$ The rotational energy levels are labeled using the conventional notation for asymmetric tops: $J_{K_{a},K_{c}}$. Numbers in parentheses represent the predicted uncertainty associated to the last digits. Transitions that were not used in the \textsc{Autofit} owing to a non-negligible contamination with a U-line are flagged with an asterisk (*).}
\end{table*}

\begin{table*}
\tabcolsep 5pt
\centering
\caption{Spectroscopic information of the transitions of $g$-ethyl formate detected toward G+0.693$-$0.027, shown in Figure \ref{f:LTEgEF}.}
\begin{tabular}{cccccccccccc}
\hline
Frequency & Transition$^{(a)}$ & log \textit{I} (300 K) & $E$$\mathrm{_{up}}$  & Blending  \\ 
(GHz) &                        &  (nm$^2$ MHz)          &  (K)                 &    \\
\hline
33.515573 (29) & 5$_{1,5}$--4$_{1,4}$  & -6.0284 & 5.1 &  Blended: U-line * \\  
35.398028 (20) & 5$_{3,3}$--4$_{3,2}$  & -6.1600 & 7.8 &  Unblended  \\        
36.604135 (25) & 5$_{1,4}$--4$_{1,3}$  & -5.9528 & 5.6 &  Blended HC$_9$N and U-line * \\  
40.121894 (74) & 6$_{1,6}$--5$_{1,5}$  & -5.7904 & 7.0 &  Blended: H$_2$NCO and U-line * \\  
40.929526 (64) & 6$_{0,6}$--5$_{0,5}$  & -5.7656 & 6.9 &  Blended: C$_2$H$_5$CHO  \\  
42.107618 (48) & 6$_{2,5}$--5$_{2,4}$  & -5.7887 & 8.3 &  Unblended  \\  
42.509261 (48) & 6$_{3,4}$--5$_{3,3}$  & -5.8557 & 9.9 &  Blended \\ 
42.622312 (49) & 6$_{3,3}$--5$_{3,2}$  & -5.8534 & 9.9 &  Blended: U-line * \\ 
43.765633 (58) & 6$_{1,5}$--5$_{1,4}$  & -5.7165 & 7.6 &  Unblended  \\  
46.69026 (16)  & 7$_{1,7}$--6$_{1,6}$  & -5.5918 & 9.2 &  Blended: CH$_3$C$_6$H and U-line *\\  
\hline 
\end{tabular}
\label{tab:gethylformate}
\vspace*{-0.5ex}
\tablefoot{$^{(a)}$ The rotational energy levels are labeled using the conventional notation for asymmetric tops: $J_{K_{a},K_{c}}$. Numbers in parentheses represent the predicted uncertainty associated to the last digits. Transitions that were not used in the \textsc{Autofit} owing to a non-negligible contamination with a U-line are flagged with an asterisk (*).}
\end{table*}

\begin{table*}
\tabcolsep 5pt
\centering
\caption{Spectroscopic information of the transitions of hydroxyacetone detected toward G+0.693$-$0.027, shown in Figure \ref{f:LTEHA}.}
\begin{tabular}{cccccccccccc}
\hline
Frequency & Transition$^{(a)}$ & log \textit{I} (300 K) & $E$$\mathrm{_{up}}$  & Blending  \\ 
(GHz) &                        &  (nm$^2$ MHz)          &  (K)                 &    \\
\hline
34.06171 (10)   &  5$_{1,5}$--4$_{0,4}$ (A) & -5.9719 & 4.7 &  Unblended  \\  
35.10788 (10)   &  6$_{0,6}$--5$_{1,5}$ (A) & -5.8670 & 6.4 &  Slightly blended: H$^{13}$CC$^{13}$CN  \\ 
36.6000352 (10) &  6$_{3,4}$--6$_{2,5}$ (A) & -5.9821 & 9.7 &  Blended: CH$_3$C(O)OCH$_3$  \\ 
36.62067 (10)   &  6$_{1,6}$--5$_{1,5}$ (A) & -5.9796 & 9.7 &  Blended: U-line *  \\ 
37.44757 (10)   &  6$_{0,6}$--5$_{0,5}$ (A) & -5.9551 & 6.4 &  Unblended  \\
38.96035 (10)   &  6$_{1,6}$--5$_{0,5}$ (A) & -5.7582 & 9.7 &  Blended: $cis$-CH$_3$OCHO \\
41.0700049 (58) &  6$_{2,4}$--5$_{2,3}$ (E) & -5.9122 &11.3 &  Slightly blended: U-line  \\
41.0755430 (35) &  6$_{2,5}$--5$_{2,4}$ (E) & -5.9617 &10.5 &  Slightly blended: U-line  \\
42.5091837 (15) &  7$_{1,7}$--6$_{1,6}$ (A) & -5.7830 & 8.5 &  Slightly blended: $g$-C$_2$H$_5$OCHO  \\
42.5817854 (29) &  7$_{1,7}$--6$_{1,6}$ (E) & -5.8027 &11.3 &  Blended: U-line * \\
43.0928962 (15) &  7$_{0,7}$--6$_{0,6}$ (A) & -5.7689 & 8.5 &  Unblended  \\
44.0219474 (21) &  7$_{1,7}$--6$_{0,6}$ (A) & -5.5687 & 8.5 &  Blended: Aa-n-C$_3$H$_7$OH  \\
44.808245  (47) &  9$_{4,5}$--9$_{3,6}$ (A) & -5.6091 &19.8 &  Unblended  \\
46.0448088 (31) &  7$_{2,6}$--6$_{2,5}$ (A) & -5.7438 &10.2 &  Unblended  \\
47.7984158 (16) &  8$_{0,8}$--7$_{1,7}$ (A) & -5.4297 &10.8 &  Unblended  \\
\hline 
\end{tabular}
\label{tab:hydroxyacetone}
\vspace*{-0.5ex}
\tablefoot{$^{(a)}$ The rotational energy levels are labeled using the conventional notation for asymmetric tops: $J_{K_{a},K_{c}}$. Numbers in parentheses represent the predicted uncertainty associated to the last digits. The A and E labels refer to the A- and E- symmetry states of hydroxyacetone, arising due to the presence of a methyl internal rotation motion. Transitions that were not used in the \textsc{Autofit} owing to a non-negligible contamination with a U-line are flagged with an asterisk (*).}
\end{table*}

\begin{table*}
\tabcolsep 5pt
\centering
\caption{Spectroscopic information of the transitions of lactaldehyde detected toward G+0.693$-$0.027, shown in Figure \ref{f:LTELA}.}
\begin{tabular}{cccccccccccc}
\hline
Frequency & Transition$^{(a)}$ & log \textit{I} (300 K) & $E$$\mathrm{_{up}}$  & Blending  \\ 
(GHz) &                        &  (nm$^2$ MHz)          &  (K)                 &    \\
\hline
 32.8030391 (7)  & 5$_{0,5}$--4$_{1,4}$ & -5.8204 & 5.0  &  Unblended \\
 34.8797382 (13) & 9$_{4,6}$--9$_{3,7}$ & -5.6662 & 19.7 &  Blended: aGg'-(CH$_2$OH)$_2$ *\\
 34.9104833 (7)  & 5$_{1,5}$--4$_{0,4}$ & -5.7505 & 5.1  &  Blended: U-line * \\
 39.6056410 (8)  & 6$_{0,6}$--5$_{1,5}$ & -5.5481 & 7.0  &  Unblended \\
 39.8461672 (16) & 9$_{5,4}$--9$_{4,5}$ & -5.5811 & 21.7 &  Slightly blended: U-line *\\
 40.7250108 (8)  & 6$_{1,6}$--5$_{0,5}$ & -5.5188 & 7.0  &  Unblended \\
 46.1743922 (9)  & 7$_{0,7}$--6$_{1,6}$ & -5.3312 & 9.2  &  Blended: $^{13}$CH$_2$CHCN \\
 85.7293506 (14) & 7$_{4,3}$--6$_{3,4}$ & -5.0545 & 13.6 &  Unblended * \\
101.649525 (50)  & 8$_{5,4}$--5$_{4,3}$ & -4.7871 & 18.4 &  Blended: C$_2$H$_3$CN * \\
103.911545 (50)  & 7$_{6,2}$--6$_{5,1}$ & -4.6700 & 18.0 &  Blended: N-CH$_3$NHCHO \\    
103.9116165 (19) & 7$_{6,1}$--6$_{5,1}$ & -5.3245 & 18.0 &  Blended: N-CH$_3$NHCHO \\     
103.9130904 (19) & 7$_{6,2}$--6$_{5,2}$ & -5.3245 & 18.0 &  Blended: N-CH$_3$NHCHO \\   
103.913219 (50)  & 7$_{6,1}$--6$_{5,2}$ & -4.6700 & 18.0 &  Blended: N-CH$_3$NHCHO \\
137.629993 (50)  & 9$_{8,2}$--8$_{7,1}$ & -4.3056 & 30.3 & Blended: CH$_3$C(O)OCH$_3$ \\
137.629993 (50)  & 9$_{8,1}$--8$_{7,1}$ & -4.9688 & 30.3 & Blended: CH$_3$C(O)OCH$_3$ \\
137.629993 (50)  & 9$_{8,2}$--8$_{7,2}$ & -4.9688 & 30.3 & Blended: CH$_3$C(O)OCH$_3$ \\
137.629993 (50)  & 9$_{8,1}$--8$_{7,2}$ & -4.3056 & 30.3 & Blended: CH$_3$C(O)OCH$_3$ \\
\hline 
\end{tabular}
\label{tab:lact}
\vspace*{-0.5ex}
\tablefoot{$^{(a)}$ The rotational energy levels are labeled using the conventional notation for asymmetric tops: $J_{K_{a},K_{c}}$. Numbers in parentheses represent the predicted uncertainty associated to the last digits. Transitions that were not used in the \textsc{Autofit} owing to a non-negligible contamination with a U-line or a lower S/N ratio are flagged with an asterisk (*).}
\end{table*}

\begin{table*}
\tabcolsep 5pt
\centering
\caption{Spectroscopic information of the transitions of methoxyacetaldehyde detected toward G+0.693$-$0.027, shown in Figure \ref{f:LTEMeta}.}
\begin{tabular}{cccccccccccc}
\hline
Frequency & Transition$^{(a)}$ & log \textit{I} (300 K) & $E$$\mathrm{_{up}}$  & Blending  \\ 
(GHz) &                        &  (nm$^2$ MHz)          &  (K)                 &    \\
\hline
  35.42961 (15)   &  8$_{1,8}$--7$_{1,7}$  & -5.1429 & 8.8  &  Unblended \\
  35.84289 (15)   &  8$_{0,8}$--7$_{0,7}$  & -5.1243 & 7.7  &  Slightly blended: $^{13}$CH$_2$CHCN  \\
  39.85602 (15)   &  9$_{1,9}$--8$_{1,8}$  & -4.9907 & 10.7 &  Unblended \\
  40.3145715 (17) &  9$_{0,9}$--8$_{0,8}$  & -4.9737 & 9.6  &  Unblended \\
  40.3896344 (17) &  9$_{2,7}$--8$_{2,6}$  & -5.0014 & 14.6 &  Slightly blended: U-line \\  
  44.2818158 (23) & 10$_{1,10}$--9$_{1,9}$  & -4.8554 & 12.9 &  Slightly blended: U-line \\
  44.7831474 (18) & 10$_{0,0}$--9$_{0,9}$  & -4.8396 & 11.7 &  Blended: CH$_3$C$_4$H \\
  44.8303900 (18) & 10$_{2,9}$--9$_{2,8}$  & -4.8637 & 16.8 &  Blended: H$^{13}$COOH \\
  44.8860560 (19) & 10$_{2,8}$--9$_{2,7}$  & -4.8626 & 16.8 &  Blended: U-line *\\
  45.3688132 (25) & 10$_{1,9}$--9$_{1,8}$  & -4.8347 & 13.1 &  Blended: C$_2$H$_5$CHO \\
  49.2483433 (20) & 11$_{0,11}$--10$_{0,10}$ & -4.7190 & 14.1 &  Blended: CH$_3$$^{34}$SH \\
\hline 
\end{tabular}
\label{tab:meta}
\vspace*{-0.5ex}
\tablefoot{$^{(a)}$ The rotational energy levels are labeled using the conventional notation for asymmetric tops: $J_{K_{a},K_{c}}$. Numbers in parentheses represent the predicted uncertainty associated to the last digits. Transitions that were not used in the \textsc{Autofit} owing to a non-negligible contamination with a U-line are flagged with an asterisk (*). }
\end{table*}

\end{appendix}

\end{document}